\documentclass[lettersize,journal]{IEEEtran} 
\usepackage[noadjust]{cite}
\usepackage{amsmath, amsfonts, amssymb}
\usepackage{algorithmic}
\usepackage{graphicx}
\usepackage{textcomp}
\usepackage{xcolor}
\usepackage{caption}
\usepackage{subcaption}
\usepackage{algorithm}
\usepackage{array}
\usepackage{stfloats}
\usepackage{url}
\usepackage{verbatim}
\usepackage{pifont}
\usepackage[normalem]{ulem}
\usepackage{makecell}
\usepackage{array}

\begin{document}

\setlength{\abovedisplayskip}{5pt}
\setlength{\belowdisplayskip}{5pt}

\title{SpikingRx: From Neural to Spiking Receiver}

\author{Ankit Gupta,~\IEEEmembership{Member,~IEEE,}, Onur Dizdar,~\IEEEmembership{Member,~IEEE,}, Yun Chen,~\IEEEmembership{Member,~IEEE,}, and \\ Stephen Wang~\IEEEmembership{Senior Member,~IEEE.}
\thanks{The authors are with VIAVI Marconi Labs, VIAVI Solutions Inc., Stevenage SG1 2AN, UK. (e-mail: ankit.gupta@viavisolutions.com; onur.dizdar@viavisolutions.com; yun.chen@viavisolutions.com; stephen.wang@viavisolutions.com.)}}


\IEEEpubid{0000--0000/00\$00.00~\copyright~2021 IEEE}

\maketitle

\begin{abstract}
In this work, we propose an energy efficient neuromorphic receiver to replace multiple signal-processing blocks at the receiver by a Spiking Neural Network (SNN) based module, called SpikingRx. We propose a deep convolutional SNN with spike-element-wise ResNet layers which takes a whole OFDM grid compliant with 5G specifications and provides soft outputs for decoded bits that can be used as log-likelihood ratios. We propose to employ the surrogate gradient descent method for training the SpikingRx and focus on its generalizability and robustness to quantization. Moreover, the interpretability of the proposed SpikingRx is studied by a comprehensive ablation study. Our extensive numerical simulations show that SpikingRx is capable of achieving significant block error rate performance gain compared to conventional 5G receivers and similar performance compared to its traditional NN-based counterparts with approximately  $9\times$ less energy consumption.
\end{abstract}
\vspace{-0.4cm}

\section{Introduction}
\IEEEPARstart{T}{he} advances in Artificial Intelligence (AI) for wireless communications have appeared as a potential enabler to achieve the requirements of the sixth generation (6G) wireless networks \cite{Dorner2018}. The Third Generation Partnership Project (3GPP) has already started to work on standardizing the use of AI in wireless communications, broadly named AI for Air Interface in \cite{3GPP38843}. The current strategy in 3GPP is to replace individual signal processing blocks for performance enhancement, with a focus on interface and terminology design. Naturally, a possible next step for the 3GPP is to replace multiple signal processing blocks for the AI for Air Interface \cite{Hoydis2021}. 

Receiver architectures employing neural networks (NNs) to replace multiple signal processing blocks, named neural receivers (NeuralRx), have appeared as a promising candidate for the future AI for Air Interface due to the attention from both academia and industry \cite{Honkala2021, Cammerer2020, Gupta2023, Faycal2022, Pihlajasalo2023, Raviv2023, Xie2024, Mei2024, Sun2024, Korpi2023, Wan23, Gupta2022, Gupta2023_2, Gupta2021_2}. NeuralRx replaces multiple signal processing blocks such as channel estimation and interpolation, equalization, and symbol de-mapping by a single NN. The motivation behind neural receivers is to deal with the channel effects and hardware impairments with higher accuracy as the joint training of the multiple blocks enables the NN to remove the inherent assumptions made during the design of the signal-processing blocks for mathematical tractability, leading to significant performance gains compared to receivers with separate signal processing blocks.

Artificial NNs (ANNs) have been the conventional building blocks to achieve NN-based processing for wireless communications. Recently, Spiking Neural Networks (SNNs) have been proposed for wireless communications applications as a low-power substitute for the ANN-based frameworks. SNNs replace the conventional neurons in ANNs with spiking neurons. Consequently, SNNs employ discrete temporal $1$-bit spikes to encode the data whereas ANNs employ real-value numbers. 
The human brain is a very sparse neural network since most neurons are normally dormant, which contributes to the remarkable energy- and space-efficiency of biological neural networks. In this sense, the SNNs resemble the human brains more than ANNs, and they can be used with effective neuromorphic computing systems \cite{Hyeryung2021, Skatchkovsky2021_2, Rajendran2019}. Recent studies have shown that by using SNNs, one can reduce the energy consumption up to $10-20$ times compared to traditional ANN architectures \cite{Ge2023, Ortiz2024, Liu2024, Liu2024_2, Chen2023, Chen2024, Borsos2022, Velusamy2023, Vogginger2022, Wen2024, Chen2023_2, Dakic2024, Xie2022, Hamedani2020, Hamedani2021}. Each spike in an SNN requires as little as several picojoules of energy, and the amount of energy consumed is practically proportional to the volume of spikes undergoing processing \cite{Rajendran2019}. Owing to their benefits, several application areas have been proposed for SNNs in wireless communications, such as Integrated Sensing and Communications (ISAC) \cite{Skatchkovsky2021_2, Vogginger2022, Wen2024, Chen2023_2}, semantic communications with ISAC \cite{Chen2023}, sum-rate maximization \cite{Ge2023}, spectrum sensing \cite{Hamedani2021, Liu2024, Hamedani2020}, satellite communications \cite{Ortiz2024, Dakic2024}, distributed wireless networks \cite{Borsos2022}, and distributed routing \cite{Velusamy2023}. 
However, the power consumption and energy-efficiency benefits of SNNs come with a cost in the form of performance degradation compared to ANNs. One reason for this is that the research on SNNs is still in its infancy, with many open problems, such as the best way to train the non-differentiable spikes, how to create a deep SNN, choosing activation functions, spike encoding, and spike reset mechanism.
\vspace{-0.34cm}
\subsection{Related Work}
Designing a NeuralRx by employing an ANN is a widely investigated topic \cite{Honkala2021, Cammerer2020, Gupta2023, Faycal2022, Pihlajasalo2023, Raviv2023, Xie2024, Mei2024, Sun2024, Korpi2023, Wan23, Gupta2022, Gupta2023_2, Gupta2021_2}. We can broadly classify the related works based on their training methodology, {\sl i.e.}, receiver-side and end-to-end (E2E). The works that focus on only the receiver side training replace the channel estimation, interpolation, equalization, and symbol de-mapping blocks at the receiver side. These works consider that the transmitted signal contains pilots. E2E training is another approach, wherein the NN blocks at the transmitter and receiver are trained jointly to achieve higher performance gains than only transmitter or receiver side training. The studies using E2E training mainly focus on pilotless OFDM transmissions. Specifically, in addition to replacing the signal processing blocks as in the receiver side training, the E2E training works also replace the signal mapping block at the transmitter, which provides custom constellation designs for the pilotless transmission \cite{Faycal2022, Cammerer2020}.

The above-mentioned works employ an ANN for designing the NeuralRx. However, employing an ANN in 6G devices and networks will exacerbate the concerns of power consumption and energy efficiency, and limit the practical applications of the NeuralRx for devices with strict power consumption requirements, such as user equipment and battery-powered Internet-of-Things (IoT) devices. 3GPP also raises this concern in~\cite{3GPP38843}, and thus focuses on use cases which require smaller AI modules or where the AI module is deployed mostly at the base station. This motivates the use of SNNs for NeuralRx architectures that can provide more ``intelligence-per-joule".

However, there are only a few works on the use of SNNs for designing the neural receivers \cite{Liu2024_2, Chen2023_2}. 
In \cite{Liu2024_2}, an SNN-based symbol detector for MIMO-OFDM systems is proposed, where the symbol detection problem is treated as a regression problem. An SNN module is employed at the receiver for channel estimation, equalization, and symbol de-mapping. The SNN is trained using a knowledge distillation-based teacher-student learning algorithm, such that an ANN-based Echo State Network is used as the teacher and SNN-based Liquid State Machine (reservoir computing) is the student. Although reservoir computing training is highly energy-efficient because training is performed only for the output layer, its learning capability and interpretability remain limited due to the fixed and random nature of the reservoir, respectively. The authors in \cite{Chen2023_2} propose neuromorphic ISAC, where an SNN is deployed at the receiver to decode digital data and detect the radar target. The transmission is performed in the form of neuromorphic communications by using impulse radio (IR) transmission and pulse position modulation (PPM) method for symbol mapping. An SNN is employed to provide two binary values as output: (i) for information bits {0, 1}, and (ii) for target detection (presence/absence). However, the authors do not consider conventional waveforms ({\sl e.g.,} OFDM) or modulation schemes ({\sl e.g.,} Quadrature Amplitude Modulation (QAM)) that are widely used in practical systems for transmission.

Although the abovementioned works investigate several aspects of the use of SNNs for neural receivers, there is still a lack of a comprehensive study of SNNs for neural receivers in terms of design, architectures, training methods, generalizability aspects, and performance improvements achieved for practical systems such as 5G New Radio (5G-NR). 
\vspace{-0.15cm}
\subsection{Contributions}
The contributions of the paper can be listed as follows:
\begin{enumerate}
    \item We propose a novel architecture for SNN-based neural receivers, called SpikingRx, which is a deep SNN module based on a convolutional neural network (CNN) and spike-element-wise residual network (SEW ResNet) to replace the channel estimation, equalization, interpolation and symbol de-mapping jointly. The proposed network takes the whole 5G-NR resource grid in a slot with different DMRS configurations as input and produces soft outputs for each resource element of the OFDM resource grid, which can be used as log-likelihood ratios (LLRs) for soft channel decoding algorithms.
    \item We treat the symbol detection problem as a multi-label classification problem in the proposed SpikingRx architecture. We combine SNN and ANN by employing the SNN in all layers except the last layer, which is designed using an ANN with a sigmoid activation function to enhance the detection performance. SEW ResNet blocks are used instead of traditional ResNet blocks to achieve identity mapping and tackle the vanishing gradient problem. 
    \item We propose the surrogate gradient descent (SGD) method for training the SpikingRx. We focus on the generalizability aspect during training in terms of signal-to-noise ratio (SNR), Doppler, delay, and channel models, such that the network can adapt to various environments without any additional training. Furthermore, we perform ``quantize-aware'' training, where the weights are quantized during the training to achieve robust performance under quantization.
    \item We perform an extensive ablation study to have an in-depth understanding of the performance of SpikingRx for communications applications. We investigate the activation probabilities, and membrane potential and observe the performance with training epochs, varying activation neurons, time-steps, surrogate functions, SEW ResNet block and its combining operations.
    \item We demonstrate that the proposed SNN-based neural receiver architecture and training method achieves a significant communications performance gain compared to conventional signal processing-based receivers. Moreover, we show for the first time that SNN-based receivers can achieve performance as good as their ANN-based counterparts with around $9\times$ less energy consumption. 
\end{enumerate}
The organization of the paper is as follows. Section II describes the system model. We describe the proposed architecture and the training method in Section III. Section IV presents the ablation study and numerical results are performed in Section V. Section VI concludes the paper.

\textit{Notation:}
Matrices and vectors are denoted by bold uppercase and lowercase letters, respectively. The
operations $|.|$ and $||.||$ denote the absolute
value of a scalar and l2-norm of a vector, respectively. $\mathcal{CN} (0, \sigma^2)$ denotes the Circularly Symmetric Complex Gaussian distribution with zero mean and variance $\sigma^2$. Logarithms are natural logarithms, {\sl i.e.}, $\log(.) = \log_{e}
(.)$, unless stated otherwise. The operator $\lfloor\cdot\rceil$ represents round-to-nearest integer operation. $\mathbb{C}$ and $\mathbb{R}$ denote the set of complex and real numbers, respectively.
\section{System Model}
\label{sec:system_model}
\begin{figure}[t!]
    \centering
    \includegraphics[scale=0.38]{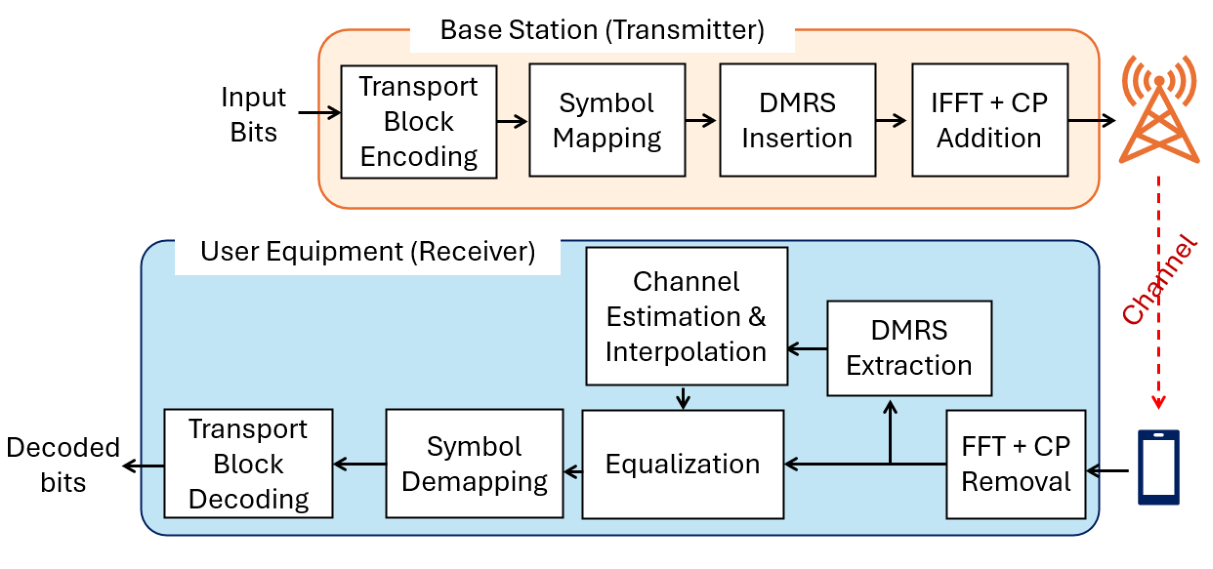}
    \caption{Block diagram of SISO downlink PDSCH transmission. The receiver consists of conventional signal-processing blocks.}
    \vspace{-0.5cm}
   \label{fig:architecture_conv}
\end{figure}

In this work, we consider a single-user downlink scenario, where the base station (BS) transmits a physical layer downlink shared channel (PDSCH) to the user equipment (UE). We consider Single-Input Single-Output (SISO) channel for the sake of simplicity, however, the SpikingRx proposed in this work remains directly extendable to multiple antennas at the BS/UE (Single-Input Multiple-Output (SIMO)/Multi-Input Single-Output (MISO) channels) without any modifications.

\subsection{PDSCH Signal Transmission-Reception}
In Fig.~\ref{fig:architecture_conv}, we show a conventional transceiver architecture for PDSCH transmission and reception. The information bits are fed into the transport block encoder, which outputs a series of codewords to be transmitted. Specifically, the input information bit sequence is divided into several parts, which are encoded using Low-Density Parity Check (LDPC) codes and applied rate-matching to obtain codewords. The final bit sequence to be transmitted is obtained by concatenating the codewords after scrambling. Next, complex baseband symbols are created using the resulting codeword sequence, which is mapped to the Physical Resource Blocks (PRBs) in a Transmission Time Interval (TTI), named as the resource grid, as shown in Fig.~\ref{fig:BW_SNN}a. Additionally, pilots, also referred to as Demodulation Reference Signal (DMRS) are injected into specifically defined subcarriers and OFDM symbols. Next, the PRBs are input to an inverse fast Fourier transform (IFFT) block, which turns complex baseband symbols into time-domain OFDM symbols. Finally, a cyclic prefix (CP) is appended at the beginning of every OFDM symbol to reduce inter-symbol interference.

The obtained signal is transmitted through the channel. The received signal at the UE is distorted by Additive White Gaussian Noise (AWGN). The UE removes the CP and applies a fast Fourier transform (FFT) on each OFDM symbol. The received signal at the output of the FFT block is expressed as
\begin{align}
y_{m,n} = h_{m,n} x_{m,n} + z_{m,n},
\label{eqn:received}
\end{align}
where $x_{m,n}\in\mathbb{C}$ and $y_{m,n}\in\mathbb{C}$ denote the transmitted and received signals, respectively, $h_{m,n}\in\mathbb{C}$ denotes the channel between the BS and UE, and $z_{m,n} \sim \mathcal{CN}(0,N_0)$ is the AWGN at the $m$-th OFDM symbol and $n$-th subcarrier for $m\in\left\{0,\ \ldots,\ \ M-1\right\}$ and $n\in\left\{0,\ldots,\ \ N-1\right\}$.
\vspace{-0.1cm}
\subsection{5G-NR Receiver Processing}
As illustrated in Fig.~\ref{fig:architecture_conv}, channel estimation is performed after FFT operation to determine the channel estimate by employing the known pilot/DMRS symbols, $p_{i,j}\in\mathbb{C}$, where $i\in \mathcal{I} \subseteq \left\{0,\ \ldots,\ \ M-1\right\}$ and $j\in\mathcal{J} \subseteq\left\{0,\ldots,\ \ N-1\right\}$ represent the OFDM symbol and subcarrier that the pilot is located in, respectively, and  $\mathcal{I}$ and $\mathcal{J}$ are the sets that contain the symbol and subcarrier indexes of DMRS symbols, respectively. We consider the least square (LS) algorithm to obtain the channel estimates. Accordingly, the channel estimate and the error variance are obtained as
\begin{align}
\hat{h}_{i,j}\hspace{-0.05cm}=\hspace{-0.05cm}y_{i,j} \dfrac{p_{i,j}^\ast}{\left\vert p_{i,j}\right\vert^2}\hspace{-0.05cm} =\hspace{-0.05cm} h_{i,j} \hspace{-0.05cm}+ \hspace{-0.05cm}\widetilde{h}_{i,j}, \
\sigma_{i,j}^2 \hspace{-0.05cm}= \hspace{-0.05cm}\mathbb{E}\left[\widetilde{h}_{i,j}\widetilde{h}_{i,j}^\ast\right] \hspace{-0.05cm}= \hspace{-0.05cm}\dfrac{N_0}{\left\vert p_{i,j}\right\vert^2},  \nonumber
\end{align}
where ${\hat{h}}_{i,j}\in\ \mathbb{C}$ and $\widetilde{h}_{i,j}\in\ \mathbb{C}$ denote the estimated channel and the estimation error at the DMRS symbols, respectively, $\sigma_{ij}^2\in\mathbb{R}$ denotes the estimation error variance. Following the channel estimation at the pilot locations, interpolation is applied to obtain the channel estimates and error variances throughout the whole resource grid. In this work, we employ linear interpolation as it is a low-complexity algorithm that achieves a good interpolation performance. 

Next, we employ the interpolated channel $\hat{h}_{m,n}\in\mathbb{C}$ to perform Linear Minimum Mean Square Error (LMMSE) equalization on each PDSCH data symbol $y_{m^\prime n}$, $m^\prime \in \left\{0,\ \ldots,\ \ M-1\right\}\setminus\mathcal{I}$, to determine the estimated data symbols $\hat{x}_{m^\prime n}\in\mathbb{C}$, as follows:
\begin{align}
\hat{x}_{m^\prime ,n} = \left((\hat{h}_{m^\prime ,n})^{*} \hat{h}_{m^\prime ,n} + \ \sigma_{m^\prime}^2\right)(\hat{h}_{m^\prime ,n})^{*}y_{m^\prime ,n.}
\end{align}
Finally, a signal de-mapper calculates the LLRs from the equalized symbols $\hat{x}_{m^\prime ,n}$. Let us denote the $l$-th bit of the symbol at $m^\prime$-th OFDM symbol and $n$-th subcarrier by $b^{l}_{m^{\prime},n}$, $l\in\{0,\ \ldots,\ B_t-1\}$. One can obtain the LLR for $b^{l}_{m^{\prime},n}$ as 
\vspace{-0.3cm}
\begin{align}
LLR_{m^\prime ,n}^l &= \log{\left(\frac{\Pr{\left(b_{m^\prime ,n}^l=1\right|{\hat{x}}_{m^\prime ,n}^l)}}{\Pr{\left(b_{m^\prime ,n}^l=0\right|{\hat{x}}_{m^\prime ,n}^l)}}\right)} \\
&\approx\ \log{\left(\frac{\sum_{c\ \in\ \mathcal{C}_{l, 1}} \exp\left(-\left|{\hat{x}}_{m^\prime ,n}^l-\ c\right|^2/\ \sigma_{m^\prime}^2\right)}{\sum_{c\ \in\ \mathcal{C}_{l,0}} \exp\left(-\left|{\hat{x}}_{m^\prime ,n}^l-\ c\right|^2/\ \sigma_{m^\prime}^2\right)}\right)}, \nonumber
\end{align}
where $B_t$ denotes the total number of bits per symbol and $\mathcal{C}_{l,b}$ denotes the set of constellation points for the $l$-th bit being equal to $b \in\{0,1\}$.
\begin{figure}[t!]
    \centering
    \begin{subfigure}{0.18\textwidth}
        \centering
        \includegraphics[width=0.9\linewidth]{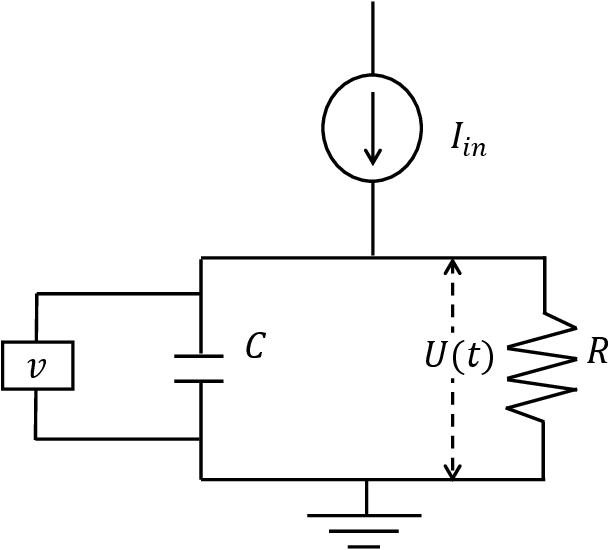}
        \vspace{1cm}
        \caption{RC circuit representation of the LIF neuron.}
        \label{fig:circuit}
    \end{subfigure}
    \begin{subfigure}{0.3\textwidth}
        \centering
        \includegraphics[width=0.97\linewidth]{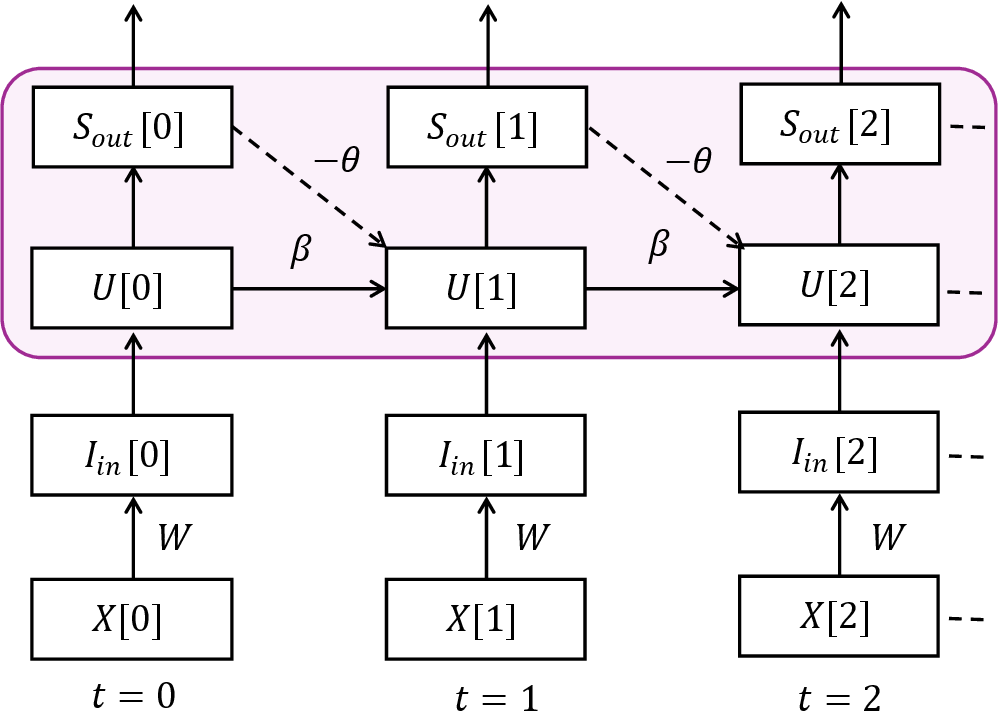}
        \caption{Computational graph of the LIF unrolled over time.}
        \label{fig:graph}
    \end{subfigure}
    \caption{Representation of the LIF neuron.}
    \vspace{-0.35cm}
\end{figure}

\section{Proposed SNN-based Receiver (SpikingRx)}
In this section, we first give background information on spiking neurons. Then, we provide details on the proposed SNN-based SpikingRx.
\vspace{-0.15cm}
\subsection{Spiking Neuron} 
\label{sec:spiking_neuron}
\begin{figure*}[t!]
    \centering
    \includegraphics[scale=0.525]{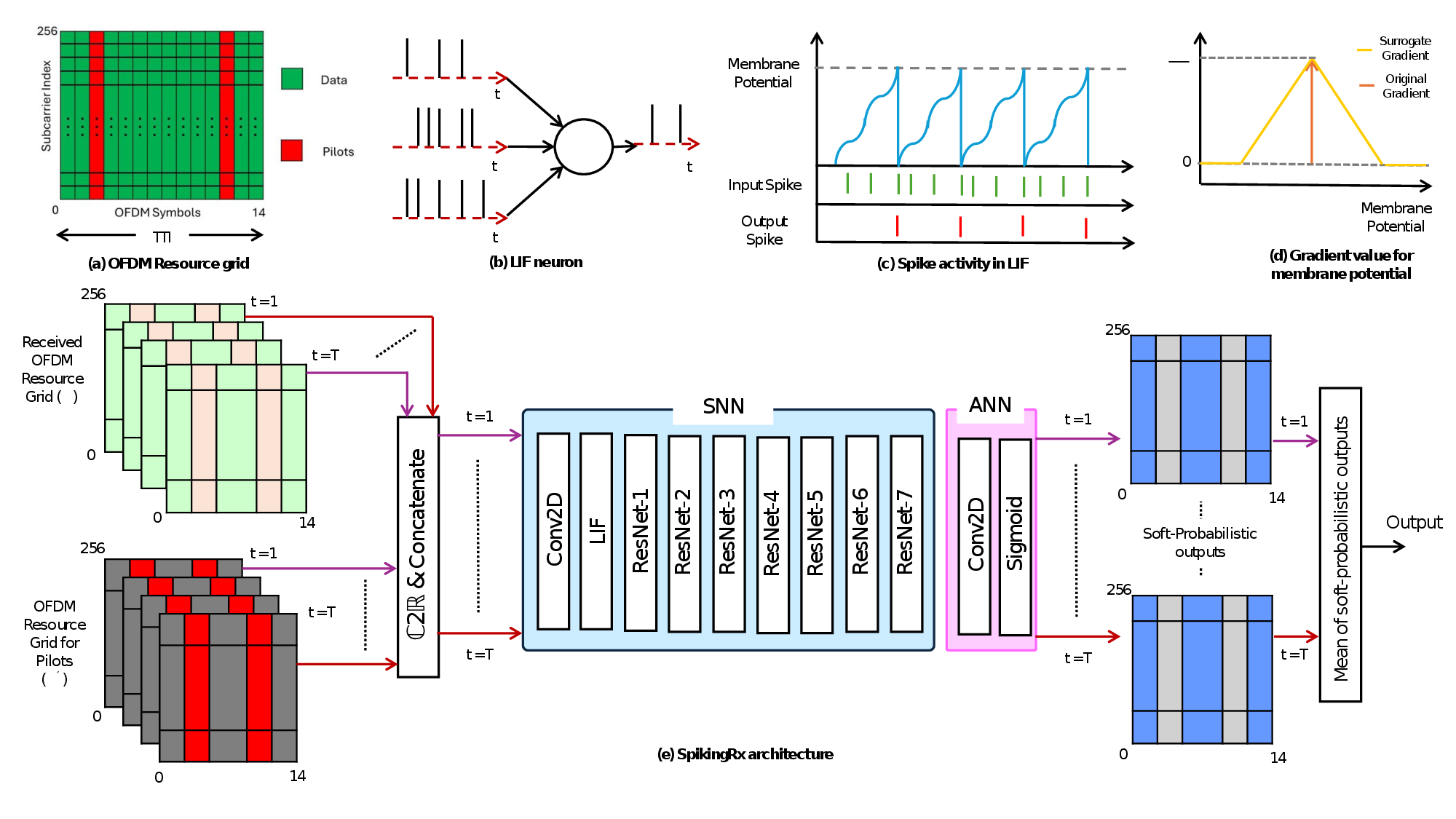}\vspace{-0.2cm}
    \caption{Illustration of SpikingRx. (a) OFDM resource grid for a TTI, (b) LIF neuron with input and output spikes, (c) Spike activities in LIF neuron, (d) Gradient values concerning membrane potential, and (e) SpikingRx architecture with concatenated received signal and pilot information as inputs, 7 ResNet blocks, and mean soft-probabilistic as output.}
    \vspace{-0.4cm}
    \label{fig:BW_SNN}
\end{figure*}

A spiking neuron is a neuron model operating on a weighted sum of inputs like an artificial neuron and is the basic building block for SNNs \cite{Eshraghian2023}. The SNNs differ from ANNs in terms of neuron activations. Specifically, while the weighted sum is passed through non-linear activation functions, such as Sigmoid, Relu, and Tanh, to obtain neuron outputs in an ANN, it simply adds to the neuron's membrane potential $U(t)$ in an SNN, so that the spiking neuron fires a spike to the following neurons only if its membrane potential crosses the threshold $\theta$. In 1907, Lapicque quantified the dynamics of SNN by showing that the spiking neuron behaves like the Leaky Integrate-and-Fire (LIF) neuron, as shown in Fig.~\ref{fig:circuit}, which is essentially a low-pass filter circuit made up of a capacitor (C) and a resistor (R)~\cite{Eshraghian2023}. By using an RC circuit, one can model the dynamics of the passive membrane as \cite{Eshraghian2023}
\begin{align}
    \tau\frac{dU(t)}{dt} = - U\left(t\right)+I_{in}\left(t\right)R,	\label{eq:du_dt}
\end{align}
where $\tau=RC$ denotes the circuit time constant. By employing Euler’s method, one can approximate the sequence-based SNN for discrete time as \cite{Eshraghian2023}
\begin{align}
    U\left[t\right]=\beta U\left[t-1\right]+\left(1-\beta\right)I_{in}[t],
    \label{eqn:u}
\end{align}
where $\beta=e^{-1/\tau}$ denotes the decay rate of $U[t]$. 
For simplicity, let us focus on the single input to single neuron scenario. By relaxing the physical viability constraint in \eqref{eqn:u}, the input current can be expressed as $I_{in}\left[t\right]=WX[t]$.
Considering the membrane potential reset and spiking as shown in Fig.~\ref{fig:graph}, we obtain the reset-by-subtraction expression as \cite{Eshraghian2023}
\begin{align}
    U\left[t\right]=\ \underbrace{\beta U\left[t-1\right]}_{\mathrm{decay}}+\ \underbrace{WX[t]}_{\mathrm{input}} - \underbrace{S_{out}[t-1]\theta.}_{\mathrm{reset}} \label{eq:u_updated}
\end{align}
where $S_{out}\in\{0,\ 1\}$ is a binary output spike, generated once the membrane potential exceeds the threshold $\theta$, as shown in Fig.~\ref{fig:BW_SNN}b, c, given by the shifted step function of Heaviside as
\begin{align}
    S_{out}\left[t\right]=\begin{cases}
        1, \qquad \text{if}\quad U[t]>\theta, \\
        0, \qquad \text{otherwise.}
    \end{cases}\label{eq:s_out}
\end{align}

\begin{figure}[t!]
    \centering
    \includegraphics[scale=0.4]{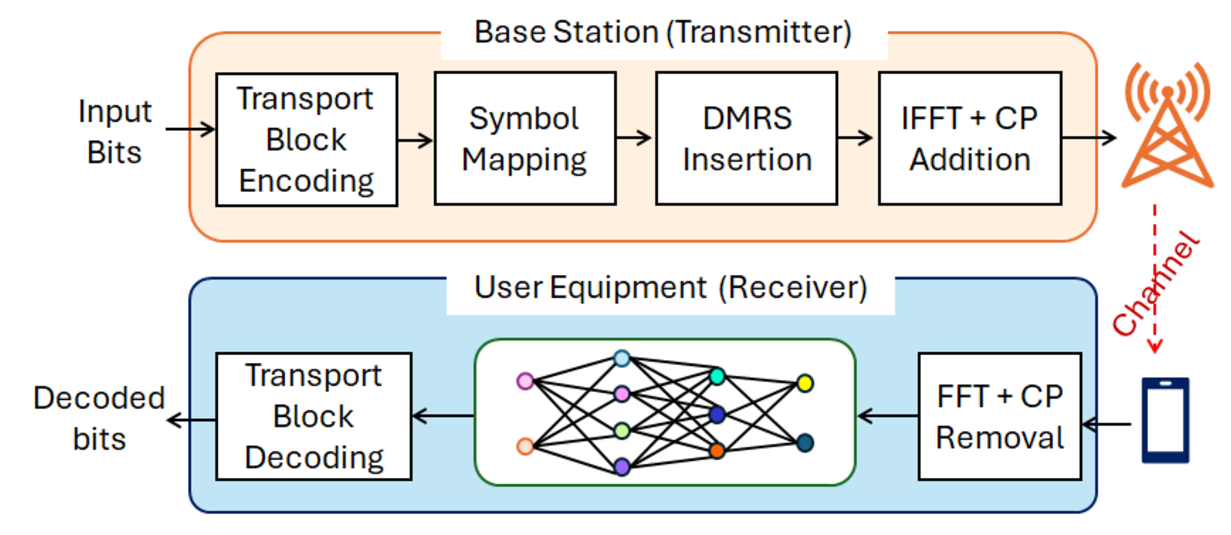}
    \caption{Block diagram of SISO downlink PDSCH transmission. The receiver consists of SpikingRx.}
    \vspace{-0.5cm}
    \label{fig:architecture_snn}
\end{figure}
\vspace{-0.3cm}
\subsection{SpikingRx Design}
In this section, we describe the architecture for SpikingRx, which is proposed to replace the signal processing blocks for channel estimation, interpolation, equalization, and symbol de-mapping, as shown in Fig.~\ref{fig:architecture_snn}. The SpikingRx is employed after the removal of CP and FFT operation. 
It takes the received frequency-domain OFDM resource grid $\mathbf{Y}\in\mathbb{C}^{M\times N}$ and the OFDM resource grid for pilots $\mathbf{P}^\prime\in\mathbb{C}^{M\times N}$ as input. The matrix $\mathbf{P}^\prime\in\mathbb{C}^{M\times N}$ contains the pilot values in corresponding pilot locations and $0$ elsewhere, as shown in Fig.~\ref{fig:BW_SNN}.
Providing the complete received resource grid with both the data and pilots enables the SpikingRx to have a better inference about the channel in both time and frequency domains\footnote{SpikingRx requires input for $T$ time-steps, thus, we propose to concatenate $T$ replications of the input before passing to the SpikingRx (see Fig.~\ref{fig:BW_SNN}).}. 

We treat the symbol detection problem as a multi-label classification problem in the proposed SpikingRx architecture. As shown in Fig.~\ref{fig:BW_SNN}e, we design the SpikingRx by concatenating the SNN layers (with spiking neurons) with the last ANN layer (with artificial neurons). This allows us to utilize Sigmoid activation in the last layer of the SpikingRx and obtain a soft output for each bit of the detected symbol in the form of probabilities. Specifically, the logits $\widehat{a_l}\in\ \mathbb{R}$ produced in the last layer of SpikingRx are passed through the Sigmoid activation function $\sigma\left(x\right) = (1+\exp{(-x)})^{-1}$ to obtain soft probabilities $\widetilde{p}(b_{m^\prime ,n}^l|\mathbf{y})$ for the $l$-th class (bits),  $l=0,\ \ldots,\ B-1$. It is shown in \cite{Cammerer2020, Gupta2021_2} that the logits correspond to the LLRs as
\begin{align}
    \widehat{a_l} = \log{\left(\frac{1-\ \widetilde{p}(b_{m^\prime ,n}^l=0|\mathbf{y})}{\ \widetilde{p}(b_{m^\prime ,n}^l=0|\mathbf{y})}\right)} = LLR_{m^\prime ,n}^l.
\end{align}
Accordingly, the soft output for $b^l_{m^{\prime},n}$ is expressed as $\widetilde{p}\left(b^l_{m^{\prime},n}=1\middle|\mathbf{y}\right)=\sigma(LLR_{m^\prime ,n}^l)$. 
SpikingRx produces $T$ soft outputs or LLRs for each RE after $T$ time steps. As shown in Fig.~\ref{fig:BW_SNN}, we calculate the mean of the outputs over the $T$ time steps to obtain the LLR or soft probability for each RE. The obtained LLRs are directly used by channel decoder\footnote{It will be shown in Section~\ref{sec:training} that the soft outputs are used in the loss function for training SpikingRx. However, the logits/LLRs can directly be taken as the SpikingRx output to be used as input to a channel decoder after the network is deployed.}. 

\textit{Remark-1}: It is possible to design SpikingRx with spiking neurons in the last layer by utilizing the rate coding. In rate coding, the predicted class is determined by the neuron that spikes most frequently. The soft-outputs $\widetilde{p}(b_{m^\prime ,n}^l|\mathbf{y})$ are obtained from the spike count by taking the mean of spikes over a total period of $T$, given as $\widetilde{p}(b_{m^\prime ,n}^l|\mathbf{y})=\sum_{t=0}^{T-1}{\vec{S_l}\left[t\right]}/T,$ where $\vec{S_l}[t]$ denote the spikes in last layer for each time step. However, in our studies, we have observed that such a design does not perform as well as the proposed design with ANN in the last layer, so we omit it for brevity.

\begin{figure}[t!]
    \centering
    \begin{subfigure}{0.5\textwidth}
        \centering
        \includegraphics[width=0.65\linewidth]{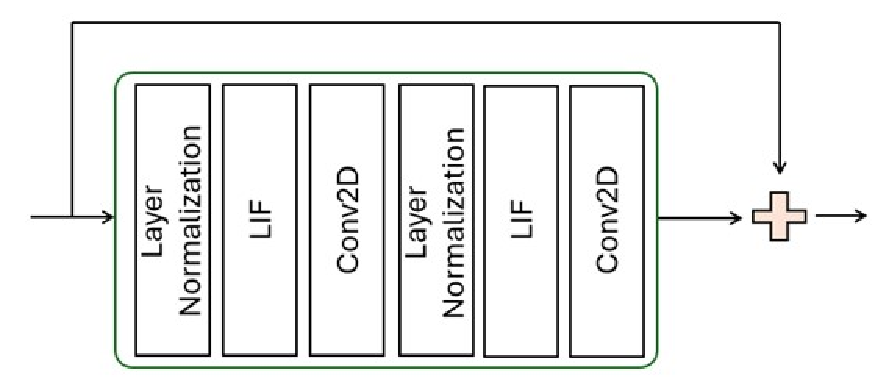}
        \caption{Traditional ResNet Block.}
        \label{fig:resnet}
    \end{subfigure}
    \begin{subfigure}{0.5\textwidth}
        \centering
        \includegraphics[width=0.6\linewidth]{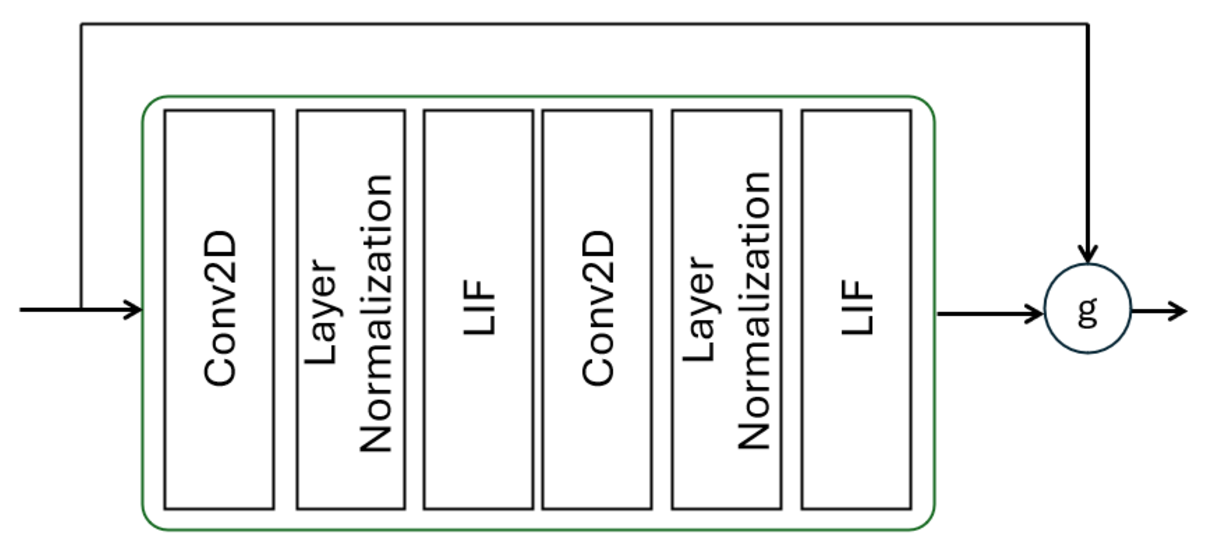}
        \caption{SEW ResNet block.}
        \label{fig:sewresnet2}
    \end{subfigure}
    \caption{Block diagrams of ResNet and SEW Resnet.}
    \vspace{-0.5cm}
\end{figure}

The detailed architecture of SpikingRx is given in Fig.~\ref{fig:BW_SNN}. We propose to utilize the 2D-convolution (Conv2D) layers so that the kernel operation is performed over the time and frequency domain to learn the channel properly. By providing the complete resource grid as input and minimizing the loss with demodulated soft bits for the whole resource grid, the SpikingRx learns channel estimation and equalization from the data symbols as well as pilot symbols. Training a deep SNN remains a challenging task due to the spiking nature of the neurons and the vanishing gradient problem, similar to the case for ANNs. 
Specifically, we can design the SpikingRx with traditional ResNet blocks by modifying the ResNet blocks proposed in \cite{Honkala2021, Cammerer2020}, such that, the ReLU activation is replaced with LIF as shown in Fig.~\ref{fig:resnet}. However, such an SNN design still suffers from two major problems~\cite{Fang2021} - (1) vanishing/exploding gradient problem (even with ResNet) because gradient of spiking neuron does not satisfy $(\partial{S}/\partial{U} = 1)$ in the SGD and (2) it is unable to achieve the identity mapping because determining a firing threshold $\theta$ that ensures $U[t]>\theta$ in \eqref{eq:s_out} is challenging.
To address the abovementioned problems, we propose to utilize the SEW ResNet block~\cite{Fang2021}, as shown in Fig.~\ref{fig:sewresnet2}. Accordingly, the first problem is tackled as by having spikes at the input and output of the ResNet block. We achieve this by changing the order of the blocks from Normalization $\rightarrow$ LIF $\rightarrow$ Conv2D to Conv2D $\rightarrow$ Normalization $\rightarrow$ LIF. The second problem is tackled by utilizing the spikes' binary properties to combine the block's input and output by various logical and element-wise operations that satisfy identity mapping (detailed in Sec.~\ref{sec:i_o_sew}). Table~\ref{table:4:snn_architecture} presents the detailed SpikingRx architecture. We implement the SpikingRx in PyTorch with the help of SNNTorch library~\cite{Eshraghian2023}.
\vspace{-0.4cm}
\subsection{Training Methodology}
\label{sec:training}
\begin{table}[t]
\caption{The SpikingRx CNN SEW ResNet.}
\label{table:4:snn_architecture}
\centering 
\renewcommand{\arraystretch}{1}
\begin{tabular}{| c | c| c |} 
\hline
\textbf{Layer} & \textbf{Filter} & \textbf{Kernel} \\ [0.5ex] 
\hline\hline
Input 1 & \multicolumn{2}{c|}{Received OFDM grid $\mathbf{Y}\in\mathbb{C}^{M\times N}$}\\
\hline
Input 2 & \multicolumn{2}{c|}{DMRS in OFDM grid $\mathbf{P}^\prime\in\mathbb{C}^{M\times N}$}\\
\hline
$\mathbb{C}2\mathbb{R}$ \& Concatenate & \multicolumn{2}{c|}{$[\mathbf{Y}, \mathbf{P}]\times T \in\mathbb{C}^{M\times N\times 2 \times T}$}\\
\hline
Conv2D & $128$ & $3\times 3$\\
\hline
LIF & \multicolumn{2}{c|}{Initial LIF}\\
\hline
Trad./SEW ResNet-1 & $128$ & $3\times 3$\\
\hline
Trad./SEW ResNet-2 & $128$ & $3\times 3$\\
\hline
Trad./SEW ResNet-3 & $128$ & $3\times 3$\\
\hline
Trad./SEW ResNet-4 & $128$ & $3\times 3$\\
\hline
Trad./SEW ResNet-5 & $128$ & $3\times 3$\\
\hline
Trad./SEW ResNet-6 & $128$ & $3\times 3$\\
\hline
Trad./SEW ResNet-7 & $128$ & $3\times 3$\\
\hline
Conv2D & $B_t$ & $1 \times 1$\\
\hline
Output & \multicolumn{2}{c|}{LLR Values $\mathbf{LLR}\in\mathbb{R}^{M^{\prime}\times N \times T}$}\\
\hline
Sigmoid  & \multicolumn{2}{c|}{Soft outputs $\widetilde{p}(b_{m^\prime ,n}^l|\mathbf{y})\in[0, 1]^{M^{\prime}\times N \times T}$}\\
\hline
Mean  & \multicolumn{2}{c|}{Soft outputs $\widetilde{p}(b_{m^\prime ,n}^l|\mathbf{y})\in[0, 1]^{M^{\prime}\times N}$}\\
\hline
\end{tabular}
\vspace{-0.4cm}
\end{table}
The designed SpikingRx solves a multi-label binary classification problem by minimizing the binary cross entropy (BCE) loss, given as 
\begin{align}
    \mathcal{L}(\cdot) &= \dfrac{1}{BMN}\sum_{l=0}^{B-1}\sum_{m=0}^{M^\prime-1}\sum_{n=0}^{N-1}b_{m^\prime ,n}^l\log\left(\widetilde{p}\left(b_{m^\prime ,n}^l\middle| \mathbf{y}\right)\right)+ \nonumber \\
    & \qquad\qquad (1-b_{m^\prime ,n}^l)\log{\left(1-\widetilde{p}\left(b_{m^\prime ,n}^l\middle| \mathbf{y}\right)\right).}
\end{align}
Any type of neural network is trained by the back-propagation method, where gradients are utilized to update the weights of the network. However, the SNN suffers from the ``dead-neuron'' problem during training \cite{Eshraghian2023}. This phenomenon occurs as the spikes are non-differentiable, such that, the gradient of the spiking neuron is zero $(\partial{S}/\partial{U} = 0)$ for all the membrane potential $(U)$ not exceeding the threshold, and $(\partial{S}/\partial{U} = \infty)$ otherwise. There are multiple methods designed to train the SNN, such as shadow training or co-learning, where an ANN is utilized to train or convert to an SNN, as done in \cite{Liu2024_2}. In this work, we propose to train the SNN from scratch without any help of a trained ANN by employing the SGD method, as shown in Fig.~\ref{fig:BW_SNN}d, which also overcomes the dead neuron problem~\cite{Friedemann2021}. Herein, the forward pass remains same as \eqref{eq:u_updated}, while during the backward pass, we approximate the non-differentiable Heaviside step-function in \eqref{eq:s_out} with a continuous differentiable function, like threshold-shifted Sigmoid function, given as $\sigma(\cdot) = (1+\exp{(\theta-U)})^{-1}$. Thus, the gradients in the backward pass are approximated as
\begin{align}
    \dfrac{\partial S}{\partial U} \rightarrow \dfrac{\partial \tilde{S}}{\partial U} = \dfrac{\exp{(\theta-U)}}{\left(\exp{(\theta-U)}+1\right)^2}.
\end{align}
Then, we can update the weights $(W)$ as
\begin{align}
    W = W - \eta \Delta_W \mathcal{L}(W),
\end{align}
where $\eta$ denotes the learning rate. In essence, the SGD enables the errors to propagate backwards irrespective of the spiking, but spiking is required to update the weights.

\textit{Remark-2:} We find that Adaptive Moment Estimation with weight decay (Adam-W) optimizer~\cite{loshchilov2019decoupledweightdecayregularization} performs better than Adam due to improved weight decay, leading to improved convergence and generalizability. 

\subsection{Generation of Training and Testing Data}
In this subsection, we detail the process of dataset creation for training and testing. We implement the 5G-NR compliant PDSCH data transmission as detailed in Sec.~\ref{sec:system_model} using Nvidia's Sionna library~\cite{sionna}. We consider an OFDM resource grid of $14$ OFDM symbols and $256$ subcarriers in a single TTI. For every TTI, an arbitrary channel model is selected. Furthermore, we randomly select the RMS delay spread, Doppler shift and SNR for every channel realization. Depending on the Doppler spread, different numbers of DMRS are required, thus we consider one DMRS and two DMRS in the OFDM resource grid. The DMRS symbols span the whole frequency grid in OFDM symbols $3$ and $12$ for two-DMRS and only $3$ for one-DMRS. Randomly generated QPSK symbols form the DMRS pilot sequences. During the training of SpikingRx, we focus on the generalizability aspect: 
\begin{itemize} 
    \item \textit{Generalizability to varying Channel Conditions} -- We consider the five different tapped delay line (TDL) channel models, each with a unique delay profile as defined by 3GPP 38.901 \cite{3GPP}. We train SpikingRx on TDL – A, C, and E and test on TDL – B, and D. 
    \item \textit{Generalizability to varying Delay Spread} -- The channel realizations for training and testing are sampled randomly for various delay spread values, ranging from $10-300$~ns. 
    \item \textit{Generalizability to varying Doppler Spread} -- We randomly sample UE velocity from $0-35$~m/s both in training and testing.
    \item \textit{Generalizability to varying SNR} –- The model can be trained for varying SNRs, by randomly sampling the SNR from $[0, 20]$~dB during training.
\end{itemize}

\begin{table}[t]
\caption{Parameters for Training and Testing.}
\label{table:5:snn_parameters}
\centering 
\renewcommand{\arraystretch}{1}
\begin{tabular}{|m{2.3cm}|m{1.5cm}|m{1.5cm}|m{1cm}|} 
\hline
\textbf{Parameter} & \textbf{Training/ Validation} & \textbf{Testing} & \textbf{Random- ization} \\ [0.5ex] 
\hline\hline
\hline
Carrier Frequency & \multicolumn{2}{c|}{$4$ GHz} & None\\
\hline
Numerology & \multicolumn{2}{c|}{$0$ ($15$ kHz subcarrier spacing)} & None\\
\hline
Number of PRBs & \multicolumn{2}{c|}{$21.33$ ($256$ subcarriers)} & None\\
\hline
Symbol Duration & \multicolumn{2}{c|}{$66.67 \mu$s} & None\\
\hline
TTI Length & \multicolumn{2}{c|}{$14$ OFDM Symbols ($1$ ms)} & None\\
\hline
Modulation Scheme & \multicolumn{2}{c|}{16-QAM} & None\\
\hline
Code-rate & \multicolumn{2}{c|}{0.5} & None\\
\hline
DMRS  & \multicolumn{2}{c|}{$1$ or $2$} & None\\
\hline
Channel Model & TDL-A, C, E & TDL-B, D & Uniform\\
\hline
$E_b/N_0$ & \multicolumn{2}{c|}{$0-20$ dB} & Uniform\\
\hline
RMS Delay Spread & \multicolumn{2}{c|}{$10-300$ ns} & Uniform\\
\hline
Max Doppler Shift & \multicolumn{2}{c|}{$0-500$ Hz} & Uniform\\
\hline
\end{tabular}
\vspace{-0.4cm}
\end{table}

\begin{figure}[t!]
    \centering
    \begin{subfigure}{0.24\textwidth}
        \centering
        \includegraphics[scale=0.55]{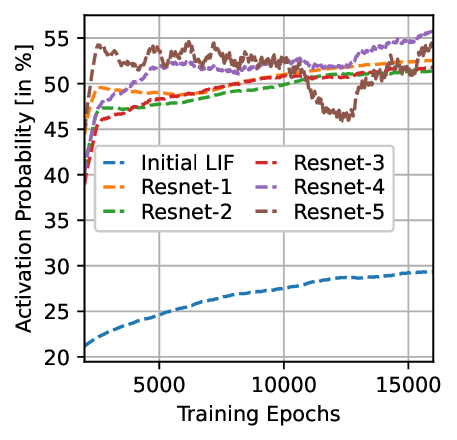}
        \caption{$5$ SEW ResNet for $T=10$.}
        \label{fig:trainact3layers}
    \end{subfigure}
    \begin{subfigure}{0.24\textwidth}
        \centering
        \includegraphics[scale=0.55]{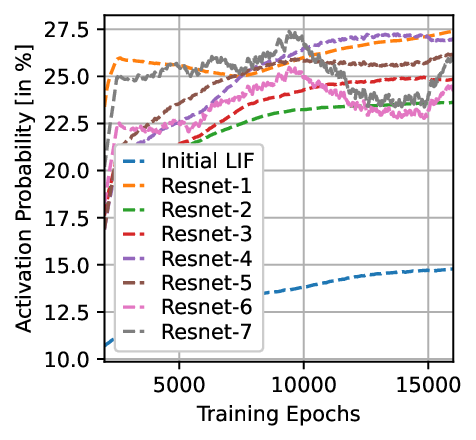}
        \caption{$7$ SEW ResNet for $T=10$.}
        \label{fig:trainact5layers}
    \end{subfigure}
    \newline
    \begin{subfigure}{0.24\textwidth}
        \centering
        \includegraphics[scale=0.55]{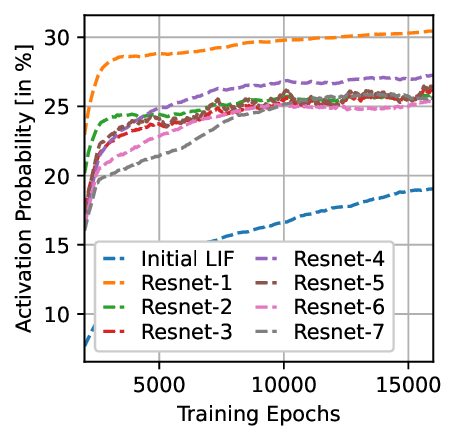}
        \caption{$7$ SEW ResNet for $T=2$.}
        \label{fig:trainact7layers}
    \end{subfigure}
    \begin{subfigure}{0.24\textwidth}
        \centering
        \includegraphics[scale=0.55]{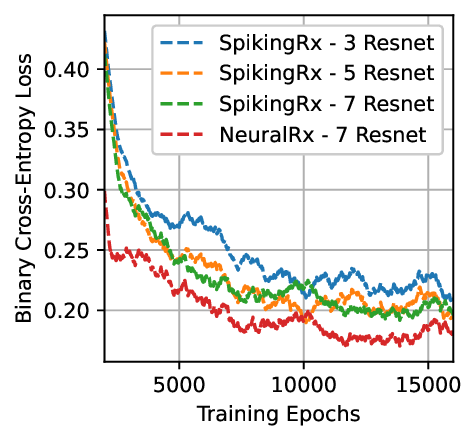}
        \caption{Validation loss for $T=2$.}
        \label{fig:trainloss357layers}
    \end{subfigure}
    \caption{Activation probability in the SpikingRx with 3/5/7 ResNet layers during the model training.}
        \vspace{-0.4cm}
    \label{fig:train_act_probability}
\end{figure}

\vspace{-0.4cm}
\subsection{Robustness of SpikingRx}
In this section, we focus on the robustness of the SpikingRx with quantized training. In our work, both ANN and SNN are trained with $32$ bits floating-point precision. However, many devices such as mobile phones, IoT devices, etc., have limited storage and processing capabilities. Further, in a wireless model update, transferring the full-precision model weights over the air will require a significant bandwidth. Thus, we focus on quantized SpikingRx.

Broadly, quantized SNN can be obtained by two methods: (1) post-training-quantization (PTQ) and (2) quantization-aware training (QAT). In PTQ, a full-precision SNN is trained and then converted to lower-precision fixed-point representations. 
In QAT~\cite{nagel2022overcoming}, the quantization of weights is performed during the forward pass in training as
\begin{align}
    \widehat{\mathbf{W}} = Q(\mathbf{W}; s, l_o, u_p) = s \cdot \text{clip}\left( \left\lfloor \dfrac{\mathbf{W}}{s} \right\rceil, l_o, u_p \right),
\end{align}
where $s$ denotes the scaling factor, and $l_o$ and $u_p$ is the lower and upper quantization threshold. However, the quantization process remains non-differentiable. Thus, we ignore it during the backward pass by using the straight-through estimator, which sets the gradients to one during the backward pass, within the quantization limits, given as
\begin{align}
    \dfrac{\partial {\mathcal{L}}}{\partial \mathbf{W}} = \dfrac{\partial {\mathcal{L}}}{\partial \widehat{\mathbf{W}}} \cdot \mathbf{1}_{l_o \leq \mathbf{w}/s \leq u_p},
\end{align}
where $\mathbf{1}$ is an indicator function that gives $1$ is $l_o \leq \mathbf{W}/s \leq u_p$ and $0$, otherwise. Thus, the model learns to tackle quantization errors during the training.
Even with further calibration, the lower-precision model obtained by PTQ suffers from significant accuracy degradation~\cite{nagel2022overcoming}. Thus, we propose to employ the QAT for obtaining the quantized SpikingRx. 

\begin{figure*}[t!]
    \centering
    \includegraphics[scale=0.40]{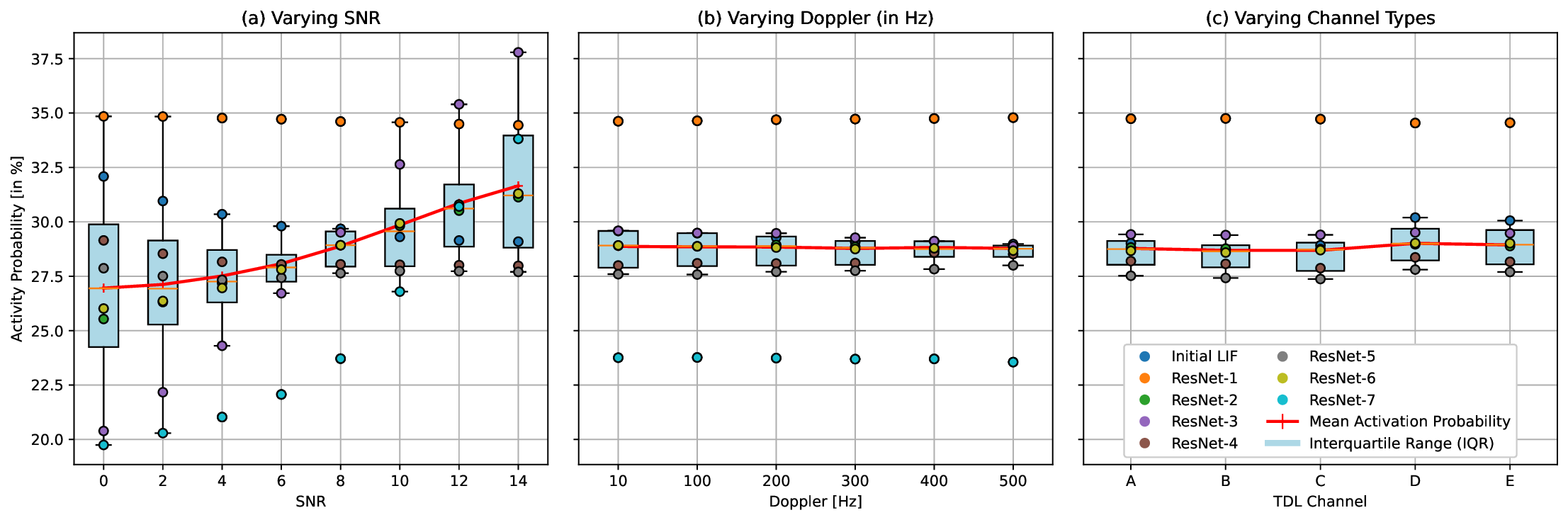}
    \caption{Activation probabilities with (a) varying SNR $(E_b/N_0)$ for low speed under testing TDL-B, D channels, (b) varying Doppler for fixed $E_b/N_0=8$~dB under testing TDL-B, D channels and (c) all five TDL channel models in \cite{3GPP}.}
    \vspace{-0.5cm}
    \label{fig:test_activation_prob}
\end{figure*}

\vspace{-0.1cm}
\section{Ablation Study}
\label{sec:ablation_study}
In this section, we perform an ablation analysis for SpikingRx to gain a more in-depth intuition of its operation. Throughout the analysis, we consider the SpikingRx architecture in Table~\ref{table:4:snn_architecture} with the parameters summarized in Table~\ref{table:5:snn_parameters}, unless stated otherwise. We train SpikingRx for $128k$ epochs with a batch size of $B=16$ and a learning rate of $\eta\leq 10^{-4}$.
\vspace{-0.2cm}
\subsection{Training Activation Probability}
The spiking neurons are said to be active when they produce discrete spikes as output at different time steps. One can calculate the spatial-temporal activation probability (in \%) of the SpikingRx as~\cite{malcolm2023}
\begin{align}
    A = \frac{100a}{BTN}, \label{eq:act_prob}
\end{align}
where $a$ denotes the number of active neurons
and $N$ is the total number of neurons. 
In Fig.~\ref{fig:train_act_probability}, we analyze the activation probability and loss convergence for varying SEW ResNet blocks and time steps ($T=2, 10$) during the training process. Figs.~\ref{fig:trainact3layers}-\ref{fig:trainact7layers} show that the activation probability of the first spiking neuron LIF on the input convolution layer is the smallest, while it remains roughly the same for all the other Resnet blocks. Intuitively, the lower activation in the first layer suggests that SpikingRx is gradually refining features as data propagates through the layers~\cite{zeiler2014visualizing}. Furthermore, similar activation in all the ResNet shows that all the blocks are learning and contributing equally to the decoded LLR output, this is because of the layer normalization in our ResNet blocks~\cite{ioffe2015batch}. 

Another important observation can be deduced from Fig.~\ref{fig:trainact3layers} and~\ref{fig:trainact5layers} by using the energy consumption calculations in Sec.~\ref{sec:energy_consumpt} is that the SpikingRx with $7$ SEW ResNet block consumes approximately $30\%$ less energy compared to one with $5$ ResNet blocks. This is in contrast to the case with ANNs, where the energy consumption increases with the increasing number of layers (of the same size). The reason for the energy consumption reduction in the SpikingRx with an increasing number of layers is that the energy consumption in SNNs also depends on spiking activity as opposed to ANNs, which reduces from approximately $50\%$ to $25\%$ as the number of ResNet blocks increases from $5$ to $7$. 

Finally, in Fig.~\ref{fig:trainloss357layers}, we investigate the BCE loss performance of SpikingRx for $T=2$ with $3$, $5$, and $7$ SEW ResNet blocks and compare their performance with that of NeuralRx. We can see that BCE validation loss values during training for all three SpikingRx designs are similar and remain close to that of NeuralRx. This implies that error performance of SpikingRx and NeuralRx are very close, in contrast to popular beliefs. 

\vspace{-0.15cm}
\subsection{Neuron Activation and Communications Performance}
In Fig.~\ref{fig:test_activation_prob}, we analyze the spiking activation probability (in \%) during the testing phase with a box plot where different layers of the SpikingRx are shown as scatter plot. We vary the SNR $(E_b/N_0)$ in  Fig.~\ref{fig:test_activation_prob}a, Doppler in Fig.~\ref{fig:test_activation_prob}b, and propagation channels in Fig.~\ref{fig:test_activation_prob}c. We can see that as the signal quality is improved the activation probability also increases. A similar observation was made in~\cite{Afshar2014} for the spike time-dependent plasticity (STDP) SNN models. Note that the improved signal quality represents the higher SNR or lower Dopper or LOS channel scenarios. 

\begin{figure}[t!]
    \centering
    \begin{subfigure}[t]{\linewidth}
        \centering
        \includegraphics[scale=0.5]{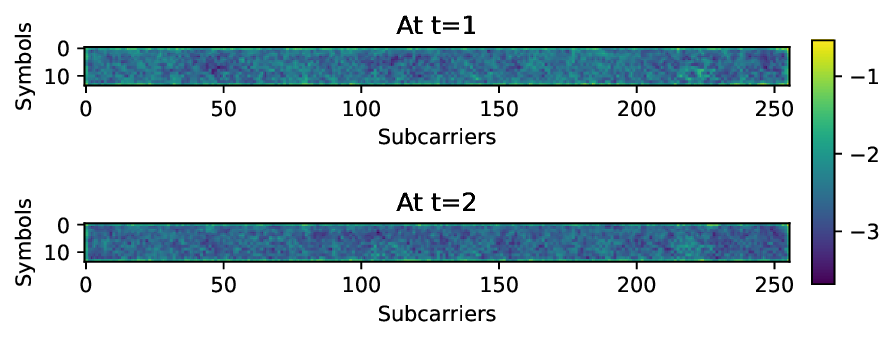}
        \caption{$E_b/N_0=0$ dB.}
        \label{fig:mem_low}
    \end{subfigure}
    \begin{subfigure}[t]{\linewidth}
        \centering
        \includegraphics[scale=0.5]{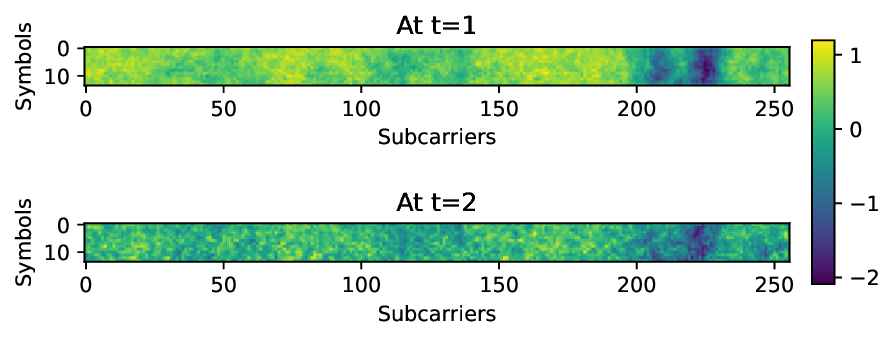}
        \caption{$E_b/N_0=14$ dB.}
        \label{fig:mem_high}
    \end{subfigure}
    \caption{Membrane potential for varying SNR for the second LIF in the SEW-ResNet-7 block.\vspace{-0.3cm}}
    \label{fig:mem_potential}
\end{figure}

To understand the above, we analyze the membrane potentials of the SpikingRx in Fig.~\ref{fig:mem_potential} for extreme SNR $(E_b/N_0)$. Specifically, we consider the membrane potential of the second LIF of the $7^\mathrm{th}$ SEW-ResNet block because it has the most significant change in its activation probability in Fig.~\ref{fig:test_activation_prob}a. Now, let us consider \eqref{eq:du_dt}, in a noisy condition such as, low SNR, high Doppler and NLOS channels, the input current $I_{in}(t)$ observes higher fluctuations because the SNR of $I_{in}(t)$ reduces. Leading to higher variability in the membrane potential $U(t)$. As detailed in Sec.~\ref{sec:spiking_neuron}, the spiking neuron fires a spike only if its membrane potential $U(t)$ crosses the threshold $\theta$, as can be seen with \eqref{eq:u_updated} and \eqref{eq:s_out}. Thus, as seen in Fig.~\ref{fig:mem_low}, for lower SNR the $U(t)\in[-3.5, -0.5]$, which is much farther from our threshold of $\theta=1$ in the SpikingRx, leading to lower activations. In contrast, for higher SNR the $U(t)\in[-2, 1]$, with most membrane potential being greater than zero, leading to higher activations. This shows that our SpikingRx is passing the most relevant information in each layer to reduce energy consumption, as discussed in Sec.~\ref{sec:energy_consumpt}.

\begin{figure}
    \centering
    \begin{subfigure}[t]{\linewidth}
        \centering
        \includegraphics[scale=0.6]{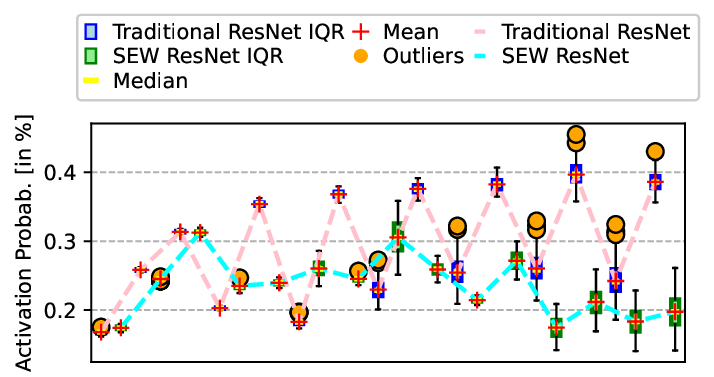}
        \includegraphics[scale=0.6]{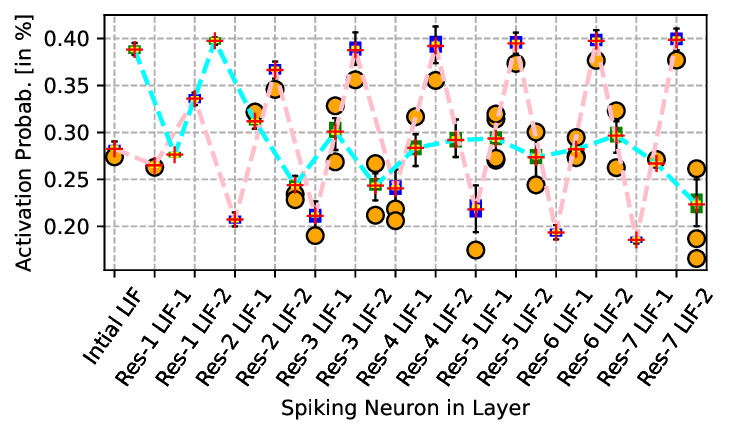}
        \caption{Activation probabilities for $T=10$ (top) and $T=2$ (bottom).}
        \label{fig:sew_res_act}
    \end{subfigure}
    \begin{subfigure}[t]{\linewidth}
        \centering
        \includegraphics[scale=0.6]{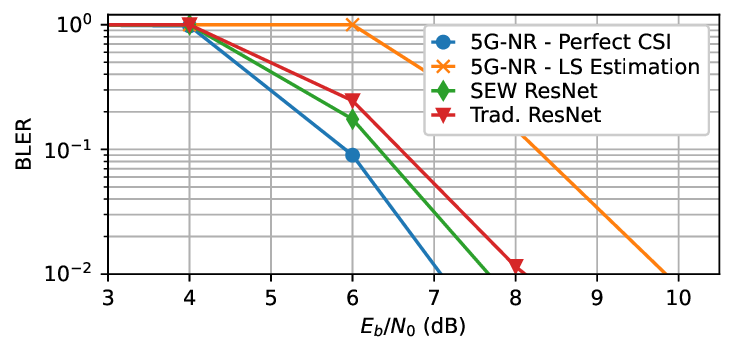}
        \caption{BLER Performance for $T=2$.}
        \vspace{-0.1cm}
        \label{fig:sew_res_bler}
    \end{subfigure}
    \caption{SEW vs. Traditional ResNet for varying time-steps.}
    \vspace{-0.4cm}
    \label{fig:sew_vs_normal_resnet}
\end{figure}

\vspace{-0.1cm}
\subsection{SEW vs. Traditional ResNet}
In Fig.~\ref{fig:sew_vs_normal_resnet} we analyze the traditional and SEW ResNet blocks for varying time steps. Fig.~\ref{fig:sew_res_act} presents the activation probabilities (in \%) for various layers. As one can see from the figure both traditional and SEW ResNet obtain mean activation probabilities of approximately $0.3$ for $T=2$. However, as the time step increases to $T=10$, the mean activation probabilities of the SEW and traditional ResNet become approximately $0.23$ and $0.35$, respectively. We use the results in Fig.~\ref{fig:sew_vs_normal_resnet} and the methods in Sec.~\ref{sec:energy_consumpt} to calculate that SEW ResNet reduces the energy consumption by $21\%$ compared to traditional ResNet for $T=10$ and remains same for $T=2$. Note that each ResNet block has two convolutions each followed by a LIF activation. Furthermore, the second LIF activates more frequently in each ResNet block, with this contrast more prevalent with traditional ResNet blocks. Intuitively, this indicates that the second convolution layer is extracting more complex features than the first convolution layer. Moreover, the activation probability of the last few ResNet blocks for SEW ResNet is much less than that of the traditional ResNet block. Thus, traditional ResNet is gradually learning with deeper layers to extract the most influential features in the last few layers, instead of spreading the learning throughout the whole SpikingRx as with the SEW ResNet block that has similar activation levels throughout. In Fig.~\ref{fig:sew_res_bler}, we analyze the BLER performance of the traditional and SEW ResNet blocks. One can see that SEW ResNet outperforms the traditional ResNet in terms of error performance.

\textit{Remark-3:} Although SEW ResNet was proposed to overcome the vanishing gradient problem in traditional ResNet for deep SNNs~\cite{Fang2021}, we do not observe such a phenomenon with either of these ResNet blocks in SpikingRx. We find that the gradients for both of them remain similar. This occurs due to the relatively small number of blocks employed in SpikingRx ({\sl e.g., $7$}) compared to the prior works using deep ResNet comprising of $50-100$ ResNet blocks~\cite{Fang2021}.

\vspace{-0.2cm}
\subsection{Input-Output Combining Operations in SEW ResNet}\label{sec:i_o_sew}
\begin{figure}[t!]
    \centering
    \includegraphics[scale=0.60]{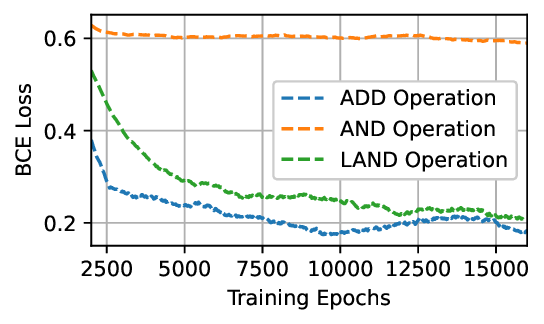}
    \caption{Operation in ResNet block.}
    \vspace{-0.4cm}
    \label{fig:various_output_oper}
\end{figure}
Since the input $(I)$ and processed signal output $(O)$ of SEW ResNet block are spiking, one can employ various logical operations to combine them as opposed to the case in traditional ResNet where a simple addition operation is used. Accordingly, one can obtain the final output of the SEW ResNet block $(g)$ using the following functions~\cite{Fang2021}:
\begin{enumerate}
    \item Addition (ADD) operation: $g=I+O$, 
    \item Logical \text{AND} operation: $g=I\ \text{AND}\ O$, 
    \item Logical IAND operation: $g=(1-I)\ \text{AND}\ O$,
\end{enumerate}
Fig.~\ref{fig:various_output_oper} shows the variations in the loss function for the abovementioned functions. It can be seen that the ordering of the functions in terms of loss convergence for the signal demapping problem is AND $<$ IAND $<$ ADD. Although not shown here for brevity, we note that the loss convergence for AND operation requires approximately $5$ times more epochs than the ones needed for other functions. This is because the gradients of SEW ADD and IAND gradually increase as they move from deeper to shallower layers due to sufficient firing rates. Furthermore, SEW ADD performs better than SEW IAND by a small margin.

\vspace{-0.2cm}
\subsection{Varying Time-Steps}
\begin{figure}[t!]
    \centering
    \begin{subfigure}[t]{\linewidth}
        \centering
        \includegraphics[scale=0.57]{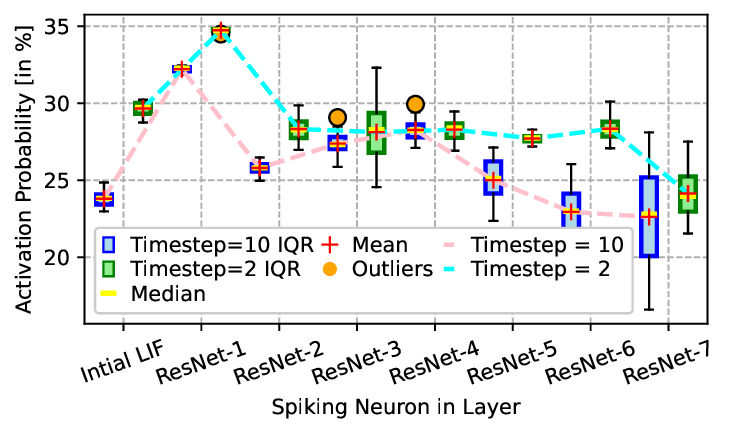}
        \caption{Activation probabilities for fixed $E_b/N_0=[0, 14]$ dB, Speed = $[0, 35]$ m/s, Channel: TDL - B,D.}
        \label{fig:timestep1}
    \end{subfigure}
    \begin{subfigure}[t]{\linewidth}
        \centering
        \includegraphics[scale=0.55]{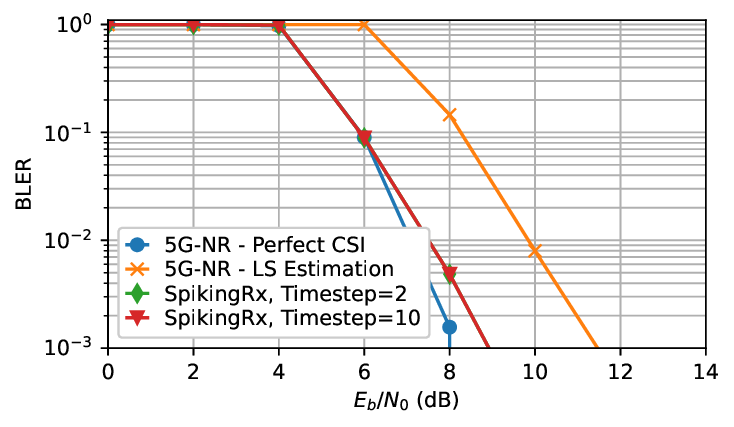}
        \caption{$E_b/N_0$ versus BLER for Speed = $10$ m/s, Channel: TDL - B,D.}
        \label{fig:timestep2}
    \end{subfigure}
    \caption{Impact of varying time-steps.}
    \vspace{-0.5cm}
    \label{fig:timesteps}
\end{figure}
In Fig.~\ref{fig:timesteps}, we analyze the impact of varying time-steps $(T)$ in the spiking LIF neurons on the SpikingRx. In Fig.~\ref{fig:timestep1}, we analyze the activation probabilities of the SpikingRx for $T=10$ and $T=2$. 
Note that the activation probability $A$ in \eqref{eq:act_prob} is inversely proportional to $T$. One can see from Fig.~\ref{fig:timestep1} that the activation probability reduces with growing time steps. To understand the effects of this phenomenon on the error performance, we analyze the BLER performance with varying time steps in Fig.~\ref{fig:timestep2}. The results demonstrate that SpikingRx achieves similar BLER performance for both $T=2$ and $T=10$. 
This indicates that the first few time steps have the maximum `relevant' information for inference, as also shown in~\cite{kim2023exploring} by analyzing temporal fisher information. The reason for this phenomenon is that we provide identical OFDM resource grids concatenated with each other as input to the SpikingRx, instead of providing rate or latency-encoded inputs as in traditional SNNs. Thus, performing larger time steps does not bring any additional benefits in terms of performance.

\begin{figure}[t!]
    \centering
    \includegraphics[scale=0.49]{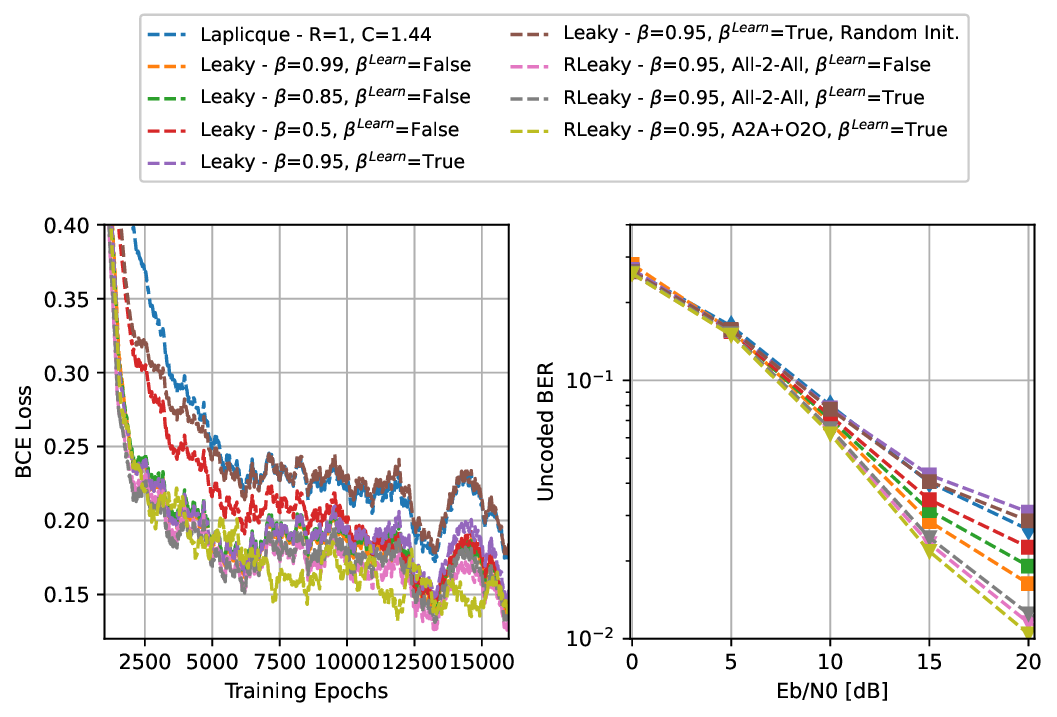}
    \caption{Varying types of spiking neuron for SpikingRx.}
    \vspace{-0.3cm}
    \label{fig:diff_activation_types}
\end{figure}
\begin{figure}[t!]
    \centering
    \includegraphics[scale=0.5]{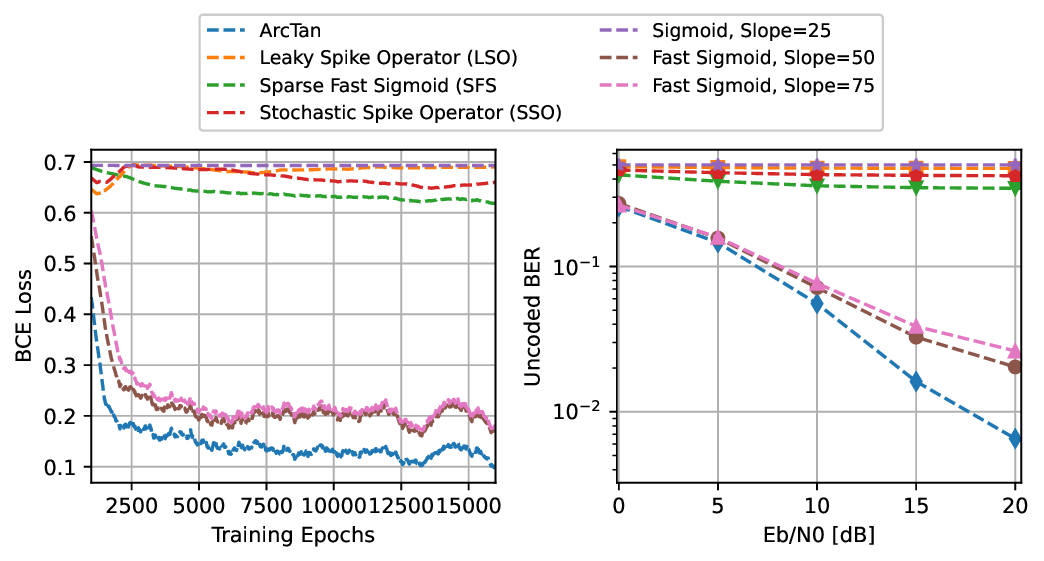}
    \caption{Varying surrogate gradients for SGD.}
    \vspace{-0.5cm}
    \label{fig:various_surrogate_gradients}
\end{figure}

\begin{figure*}[t!]
    \centering\hspace{-0.75cm}
    \begin{subfigure}[t]{.32\textwidth}
        \centering
        \includegraphics[scale=0.50]{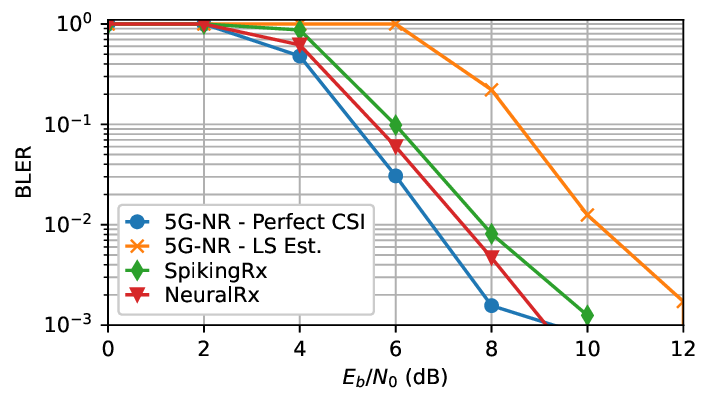}
        \vspace{-0.2cm}
        \caption{One DMRS - Low speed = 3.6 km/h.}
        \label{fig:1p_low_speed}
    \end{subfigure}
    \begin{subfigure}[t]{.32\textwidth}
        \centering
        \includegraphics[scale=0.50]{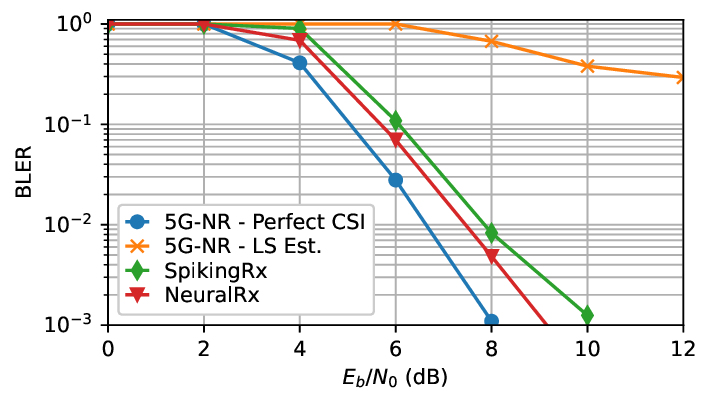}
        \vspace{-0.2cm}
        \caption{One DMRS - Med. speed = 36 km/h.}
        \label{fig:1p_med_speed}
    \end{subfigure}
    \begin{subfigure}[t]{.32\textwidth}
        \centering
        \includegraphics[scale=0.50]{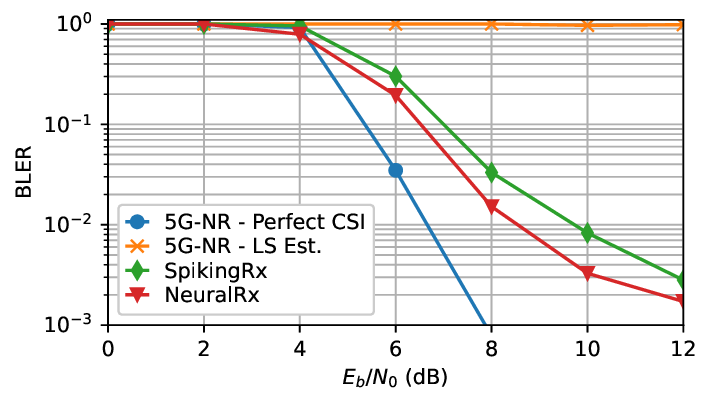}
         \vspace{-0.2cm}
       \caption{One DMRS - High speed = 108 km/h.}
        \label{fig:1p_high_speed}
    \end{subfigure}\\\hspace*{-0.5cm}
    \begin{subfigure}[t]{.32\textwidth}
        \centering
        \includegraphics[scale=0.50]{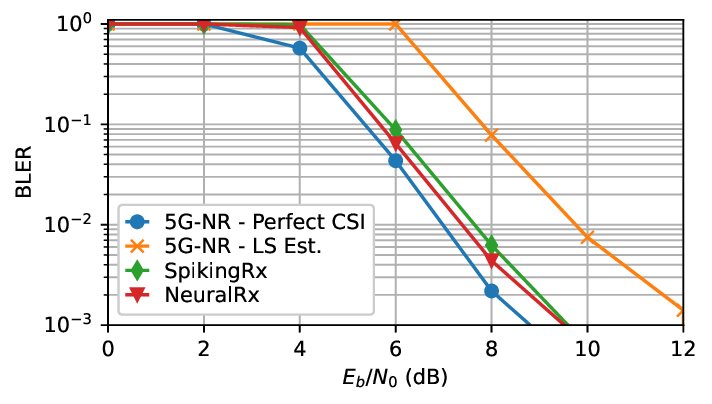}
        \vspace{-0.2cm}
        \caption{Two DMRS - Low speed = 3.6 km/h.}
        \label{fig:2p_low_speed}
    \end{subfigure}
    \begin{subfigure}[t]{.32\textwidth}
        \centering
        \includegraphics[scale=0.50]{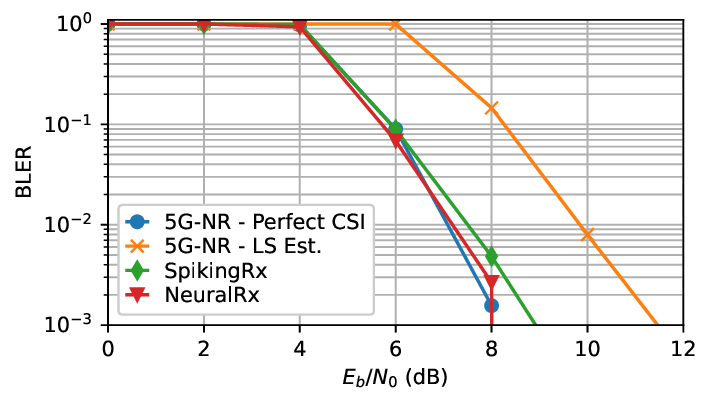}
        \vspace{-0.2cm}
        \caption{Two DMRS - Med. speed = 36 km/h.}
        \label{fig:2p_med_speed}
    \end{subfigure}
    \begin{subfigure}[t]{.32\textwidth}
        \centering
        \includegraphics[scale=0.50]{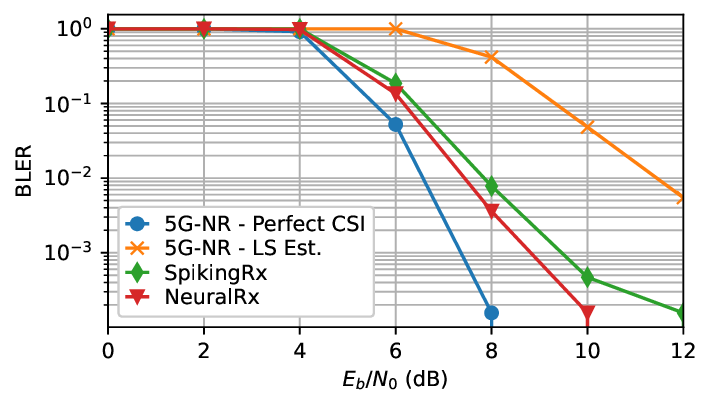}
        \vspace{-0.2cm}
        \caption{Two DMRS - High speed = 108 km/h.}
        \label{fig:2p_high_speed}
    \end{subfigure}
    \caption{BLER performance of different types of receivers for varying UE speed and pilots in OFDM resource grid.}
    \vspace{-0.5cm}
    \label{fig:bler_low_med_high}
\end{figure*}

\vspace{-0.3cm}
\subsection{Spiking Neuron Activation}
In Fig.~\ref{fig:diff_activation_types}, we analyze SpikingRx with varying spiking neurons, as detailed below~\cite{Eshraghian2023}:
\begin{itemize}
    \item Leaky -- As detailed in Sec.~\ref{sec:spiking_neuron}. 
    \item Lapicque -- Lapicque is qualitatively similar to the Leaky, except it requires the hyper-parameter setting of RC circuit parameters to determine decay rate $(\beta)$. 
    \item Recurrent Leaky -- Herein, the output spikes of the neuron is looped back to its input. RLeaky can be applied in two ways on the output spikes $S_{out}$ before returning back to input; (1) All-2-All recurrence, performing recurrent convolution operation, or (2) One-2-One recurrence, performing element-wise multiplication with $V$. 
\end{itemize}
We train SpikingRx using Fast-Sigmoid with a slope of $25$. Firstly, Fig.~\ref{fig:diff_activation_types} shows that the SpikingRx with Lapicque neurons performs the worst, since the decay rate $(\beta=e^{-1/RC})$ is calculated using the $(RC)$ hyper-parameters that require additional hyper-parameter optimization, by methods like grid-search, which was not performed here. 

Secondly, for leaky activation, we analyze the learnable and fixed decay rates. In the fixed decay, the decay rate is treated as a hyper-parameter whereas in learnable decay, the decay rate is considered as a learnable parameter. It can be observed that the learnable decay rate has worse performance due to the introduction of additional parameters and the variability in decay rates for each neuron causes instability in the neuron dynamics while training the SpikingRx. On the other hand, higher decay rates, such as $0.95-0.99$, result in enhanced performance. This can be verified from \eqref{eqn:u}, wherein a very large $\beta$ removes the dependency on the input current since $(1-\beta) I_{in} \rightarrow 0$, thereby resulting in $U[t]\approx U[t-1]$ and making the membrane potential for each time-step highly dependent on the previous time-step. Accordingly, SpikingRx makes meaningful and strong dependencies with varying time steps.

Thirdly, one can observe that the recurrent leaky neuron achieves the best performance. Although there are no temporal dependencies in our input, the recurrent connection helps to improve performance over that achieved using the normal leaky by helping to stabilize the internal representations, thus making the SpikingRx more robust to noise and random fluctuations of the wireless channel. Furthermore, recurrent leaky performs better with convolution operation. Intuitively, the channel estimation and interpolation, performed over multiple resource elements in the OFDM resource grid, is useful as it helps in capturing channel Doppler properly. In short, the neuron types can be ordered in terms of their performance in signal demapping problems as Lapicque $<$ Leaky $<$ RLeaky. 

\vspace{-0.3cm}
\subsection{Surrogate Gradient Descent}
There are multiple ways to perform surrogate gradients for the SGD. We analyze various methods to perform SGD~\cite{Eshraghian2023, Friedemann2021}: Sigmoid, Fast Sigmoid, Sparse Fast Sigmoid (SFS), Arc-tangent (ArcTan), Stochastic Spike Operator (SSO), and Leaky Spike Operator (LSO). Fig.~\ref{fig:various_surrogate_gradients} shows that only Fast Sigmoid and ArcTan surrogate gradients enable learning in SpikingRx. This occurs because the Fast Sigmoid gradient can capture the sudden shifts in membrane potential related to spike timing in neurons better than the Sigmoid gradient since they usually have sharper gradients with quicker shifts around the origin. This leads to faster convergence, robustness to noise and the effects of wireless channels, and remains more biologically plausible. Moreover, the ArcTan surrogate gradient achieves a significant performance gain over Fast Sigmoid, because ArcTan provides a smoother and continuous approximation to the gradients in the back-propagation, which helps in overcoming the problem of gradient saturation observed in Fast Sigmoid, and thus, leads to precise timing of spike and more robustness towards channel/noise impairments. The activation functions can be ordered in terms of their performance in signal demapping problems as SSO $\approx$ LSO $\approx$ SFS $\approx$ Sigmoid $<$ Fast Sigmoid $<$ ArcTan. 

\textit{Remark-4}: Note that using RLeaky instead of Leaky with ArcTan SGD provided no performance gains, despite increased complexity due to recurrent connection. Thus, the benefits of RLeaky are overshadowed by ArcTan SGD.

\section{Performance Evaluation}
\label{sec:performace_eval}
In this section, we perform further performance evaluation for SpkingRx and compare its performance with NeuralRx. Parameters are given in Table~\ref{table:5:snn_parameters} as in Section~\ref{sec:ablation_study}. We adopt the following baselines:
\begin{itemize}
    \item \textit{5G-NR - LS Estimation}: As detailed in Sec.~\ref{sec:system_model}, we consider LS channel estimation, linear channel interpolation and LMMSE equalizer. 
    \item \textit{5G-NR - Perfect Channel State Information (CSI)}: Receiver knows the perfect CSI knowledge and performs LMMSE equalization. It acts as lower bound.
    \item \textit{NeuralRx}:  We implement ANN-based counterpart of SpikingRx. Specifically, we replace SNN-based LIF neurons with ANN-based ReLU neurons and traditional ResNet blocks in SpikingRx, with the same training and testing procedure. 
\end{itemize}

\vspace{-0.3cm}
\subsection{Evaluation of SpikingRx}
In Fig.~\ref{fig:bler_low_med_high}, we evaluate the BLER performance of the SpikingRx with varying speeds and pilots. Specifically, we consider three speeds: low speed ($3.6$ km/h), medium speed ($36$ km/h), and high speed ($108$ km/h), encountered in typical walking, cycling and car scenarios, respectively. In Fig.~\ref{fig:1p_low_speed}-\ref{fig:1p_high_speed}, we consider $1$ DMRS symbol. We can observe that only NeuralRx and SpikingRx achieve BLER within $1.5$~dB of the lower bound of perfect CSI. Furthermore, the 5G-NR receiver is unable to decode the signal even for medium speed as it cannot track the channel variations using a single DMRS symbol. Another important observation is that SpikingRx achieves a performance only $0.1$ worse than that of NeuralRx, showing the potential of SNN in such applications. In Fig.~\ref{fig:2p_low_speed}-\ref{fig:2p_high_speed}, we provide the peformance results for $2$ DMRS symbols. Unlike the previous case, the 5G-NR receiver can decode the signal properly due to increased pilot density. However, it still suffers from performance degradation in the high-speed scenario. On the other hand, NeuralRx and SpikingRx achieve BLER performance within $0.2$~dB of the lower bound. 

\vspace{-0.3cm}
\subsection{Robustness of the SpikingRx}
\begin{figure}[t!]
    \centering
        \includegraphics[scale=0.60]{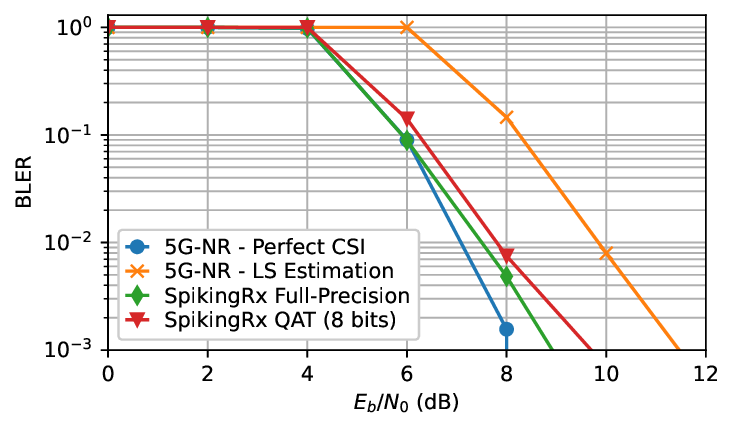}
    \caption{Robustness of SpikingRx.}
    \vspace{-0.4cm}
    \label{fig:robustSNN}
\end{figure}
In Fig.~\ref{fig:robustSNN}, we analyze the robustness of the SpikingRx to quantization. The results show that SpikingRx with $8$-bit quantized weights achieves only $0.1$~dB worse performance than that of full-precision SpikingRx with $32$ bits. Accordingly, one can conclude that SpikingRx with quantize-aware training (QAT) achieves a significant complexity gain with a negligible error rate performance degradation.

\vspace{-0.3cm}
\subsection{Energy Consumption}
\label{sec:energy_consumpt}
Finally, we compare the computational energy cost for the SpikingRx and NeuralRx. The total number of floating point operations (FLOPS) determines the entire processing overhead, which roughly corresponds to the quantity of Matrix-Vector Multiplication (MVM) operations~\cite{Xie2022}. One can calculate the number of FLOPS for the $l$-th layer in the ANN for $l=\{1, ..., L\}$ as 
\begin{align}
\text{FLOPS}_{\text{NeuralRx}}(l)\hspace{-0.1cm} = \hspace{-0.1cm} 
    \begin{cases}
         K^2 C_{in}  H_{out}  W_{out}  C_{out}, \  \text{if}\; l :=\;\text{Conv2D}, \nonumber\\
        C_{in}  H_{out}  W_{out},
        \  \text{if}\; l :=\;\text{Normalization}, 
    \end{cases}
\end{align}
where $K$ denotes the height and width of the kernel, $H_{out}$ and $W_{out}$ represent the height and width of the output, and $C_{in}$ and $C_{out}$ indicate the input and output channel. 
For the $l$-th layer in the SNN, the average firing rate per neuron or Spiking rate is given as
\begin{align}
    R_s(l) = {\sum\nolimits_{t=1}^T N_s^t(l)}\Big/{N_n(l)}, \quad \forall\;l<L
\end{align}
where $N_s^t$ and $N_n$ denote the number of spikes of the $l$-th layer for all the time steps $T$ and number of spiking neurons in that layer, respectively. Note that the last layer-L in SpikingRx is an ANN layer. For the SNN-based SpikingRx, the FLOP only happens if a spike is emitted. Thus, we can obtain the FLOP count for the $l$-th layer in the SNN as 
\begin{align}
    \text{FLOPS}_{\text{SpikingRx}}(l) = \text{FLOPS}_{\text{NeuralRx}}(l) R_s(l), \quad \forall\;l<L
\end{align}
For each input, the summation of a neuron's weighted inputs in an ANN requires a multiply and accumulate (MAC) operation, while spikes in SNNs require an accumulation (AC) operation. Thus, we can obtain the energy consumed by the $l$-th layer as 
\begin{align}
    E_{\text{NeuralRx}}(l) &= \text{FLOPS}_{\text{NeuralRx}}(l) E_{MAC}, \\
    E_{\text{SpikingRx}}(l) &= 
    \begin{cases}
        \text{FLOPS}_{\text{SpikingRx}}(l) E_{AC}, \; \forall\;l<L \\
        \text{FLOPS}_{\text{NeuralRx}}(l) E_{MAC}, \; \forall\;l=L
    \end{cases} 
\end{align}
Considering $45$nm CMOS technology, we can calculate the $E_{MAC}$ and $E_{AC}$ as in Table~\ref{table:3:energy_consump_values}~\cite{Horowitz2014}. For $Q$-bit quantization, we consider $E_{MAC}\propto Q^{1.25}$ and $E_{AC}\propto Q$~\cite{Datta2021}. Accordingly, Table~\ref{table:energy_consump_results} summarizes the energy consumption comparison. The results show that SpikingRx achieves $2.1\times$ and $8.9\times$ reduced energy consumption compared to NeuralRx for $T=10$ and $T=2$ time steps, respectively. Further, quantized SpikingRx can provide an additional $5\times$ reduced energy consumption.  

\begin{table}[t!]
\caption{Energy consumption for varying operations~\cite{Horowitz2014}.}
\label{table:3:energy_consump_values}
\centering 
\renewcommand{\arraystretch}{1}
\begin{tabular}{| c | c |} 
 \hline
 \textbf{Operation} & \textbf{Energy (pJ)} \\ [0.5ex] 
 \hline\hline
 32 bit FP MULT $(E_{MULT})$ & 3.7 \\
 32 bit FP ADD $(E_{ADD})$ & 0.9 \\
 32 bit FP MAC $(E_{MAC} = E_{MULT}+E_{ADD})$ & 4.6 \\
 32 bit FP AC $(E_{AC})$ & 0.9 \\
 8 bit FP MAC $(E_{MAC})$ & 1.1 \\
 8 bit FP AC $(E_{AC})$ & 0.2 \\
 \hline
\end{tabular}
\end{table}

\begin{table}[t!]
\caption{Energy consumption for NeuralRx and SpikingRx.}
\label{table:energy_consump_results}
\centering 
\renewcommand{\arraystretch}{1}
\begin{tabular}{|m{3.2cm}|m{1.3cm}|m{0.85cm}|m{1.35cm}|} 
 \hline
 \textbf{Neural Network} & \textbf{Energy (J)} & \textbf{Gains w.r.t. ANN}  & \textbf{Gains w.r.t. SNN (T=2)} \\ [0.5ex] 
 \hline\hline
 NeuralRx & $0.136$ & $-$  & $-$ \\
 SpikingRx $(T=10)$ & $0.071$ & $1.9\times$ & $-$ \\
 SpikingRx $(T=2)$ & $0.015$ & $8.9\times$ & $-$ \\
 Quant. SpikingRx $(T=2)$ & $0.003$ & $-$ & $5\times$  \\
 \hline
\end{tabular}\vspace{-0.35cm}
\end{table}

\section{Conclusion}
In this work, we propose an energy-efficient deep neuromorphic receiver, named SpikingRx, for symbol detection for any 5G-NR-compliant OFDM resource grid. SpikingRx is designed using deep CNN and SEW ResNet blocks concatenated with an artificial neuron layer for obtaining soft-output or LLR values for the received bits. We propose SGD to train the network. We focus on the generalizability of the SpikingRx for varying channel conditions, such as SNR, delay, and Doppler. We perform a detailed investigation on the interpretability of the SpikingRx for wireless communications. 
By numerical results, we show that SpikingRx achieves a BLER performance within $1.5$~dB and $0.2$~dB of that achieved by a receiver perfect CSI knowledge using 1 and 2 DMRS symbols, respectively, and similar to that achieved by ANN-based NeuralRx with approximately $9\times$ reduced energy consumption. 

\bibliographystyle{IEEEtran}
\bibliography{main}

\end{document}